\documentclass[numberedappendix,appendixfloats,12pt,preprint]{emulateapj}

\usepackage[]{amsmath}

\usepackage{color}			      
\definecolor{midgray}{gray}{0.4}		
\definecolor{orange}{rgb}{1,0.5,0}    

\usepackage[colorlinks=true,citecolor=midgray,linkcolor=midgray]{hyperref}        

\newcommand{\lya}{Ly$\alpha$}

\newcommand{\HST}{\textit{HST}}
\newcommand{\RXJ}{RXCJ2248.7-4431}

\shorttitle{GLASS UV Emission from $z=6.11$ Lensed Object}
\shortauthors{Schmidt et al. (2017)}

\begin{document}


\title{The Grism Lens-Amplified Survey from Space (GLASS). XI. Detection of CIV in Multiple Images of $\lowercase{z}=6.11$ \lya\ Emitter Behind \RXJ }


\author{
K.~B.~Schmidt$^{1,\star}$
K.-H., Huang$^{2}$,
T.~Treu$^{3,\ast}$,
A.~Hoag$^{2}$,
M.~Brada\v{c}$^{2}$,
A.~L.~Henry$^{4}$,
T.~A.~Jones$^{2,\dagger}$,
C.~Mason$^{3}$,
M.~Malkan$^{3}$,
T.~Morishita$^{3,5,6}$,
L.~Pentericci$^{7}$,
M.~Trenti$^{8}$,
B.~Vulcani$^{8}$,
X.~Wang$^{3}$
}
\affil{$^{1}$Leibniz-Institut f\"{u}r Astrophysik Potsdam (AIP), An der Sternwarte 16, 14482, Potsdam, Germany}
\affil{$^{2}$Department of Physics, University of California, Davis, CA, 95616, USA}
\affil{$^{3}$Department of Physics and Astronomy, UCLA, Los Angeles, CA, 90095-1547, USA}
\affil{$^{4}$Space Telescope Science Institute, 3700 San Martin Drive, Baltimore, MD, 21218, USA}
\affil{$^{5}$Astronomical Institute, Tohoku University, Aramaki, Aoba, Sendai 980-8578, Japan}
\affil{$^{6}$Institute for International Advanced Research and Education, Tohoku University, Aramaki, Aoba, Sendai 980-8578, Japan}
\affil{$^{7}$INAF - Osservatorio Astronomico di Roma Via Frascati 33 - 00040 Monte Porzio Catone, I}
\affil{$^{8}$School of Physics, The University of Melbourne, VIC, 3010 Australia}
\altaffiltext{$\star$}{\href{mailto:kbschmidt@aip.de}{kbschmidt@aip.de}}
\altaffiltext{$\ast$}{Packard Fellow}
\altaffiltext{$^\dagger$}{Hubble Fellow}

\begin{abstract}
The CIII] and CIV rest-frame UV emission lines are powerful probes of the ionizations states of galaxies. 
They have furthermore been suggested as alternatives for spectroscopic redshift confirmation of objects at the epoch of reionization ($z>6$), where the most frequently used redshift indicator, Ly$\alpha$, is attenuated by the high fraction of neutral hydrogen in the inter-galactic medium.
However, currently only very few confirmations of carbon UV lines at these high redshifts exist, making it challenging to quantify these claims.
Here, we present the detection of CIV$\lambda\lambda$1548,1551\AA\ in \HST\ slitless grism spectroscopy obtained by GLASS of a Ly$\alpha$ emitter at $z=6.11$ multiply imaged by the massive foreground galaxy cluster \RXJ.
The CIV emission is detected at the 3--5$\sigma$ level in two images of the source, with marginal detection in two other images.
We do not detect significant CIII]$\lambda\lambda$1907,1909\AA\ emission implying an equivalent width EW$_\textrm{CIII]}<20$\AA\ (1$\sigma$) and $\textrm{CIV/CIII}>0.7$ (2$\sigma$).
Combined with limits on the rest-frame UV flux from the HeII$\lambda$1640\AA\ emission line and the OIII]$\lambda\lambda$1661,1666\AA\ doublet, we put constraints on the metallicity and the ionization state of the galaxy.
The estimated line ratios and equivalent widths do not support a scenario where an AGN is responsible for ionizing the carbon atoms. 
SED fits including nebular emission lines imply a source with a mass of log(M/M$_\odot)\sim9$, SFR of around 10M$_\odot$/yr, and a young stellar population $<50$Myr old.
The source shows a stronger ionizing radiation field than objects with detected CIV emission at $z<2$ and adds to the growing sample of low-mass (log(M/M$_\odot)\lesssim9$) galaxies at the epoch of reionization with strong radiation fields from star formation.
\end{abstract}



\section{Introduction}
\label{sec:intro}
Spectroscopically confirming galaxies at the Epoch of Reionization (EoR) at redshifts above 6 has been challenging owing to the fact that Lyman-$\alpha$$\;\lambda$1216\AA\ (henceforth referred to as \lya) photons are attenuated by neutral hydrogen in the inter-galactic medium
\citep[IGM; e.g., ][]{2010ApJ...725L.205F,2011ApJ...743..132P,2011ApJ...728L...2S,2012ApJ...747...27T,2013ApJ...775L..29T,2012MNRAS.427.3055C,2014MNRAS.443.2831C}.
Dedicated efforts and improved observational tools keep providing larger and larger samples of \lya\ emitters (LAEs) at these extreme redshifts \citep[e.g.,][Hoag et al. in prep.]{Schenker:2012gn,Finkelstein:2013fx,2015ApJ...810L..12Z,2015ApJ...804L..30O,2016ApJ...818...38S,2016arXiv160202160S,2016ApJ...823L..14H,2016ApJ...817...11H,2016ApJ...823..143R}.
However, at the highest redshifts, where a significant fraction of the hydrogen in the inter-galactic medium (IGM) is neutral alternative methods of redshift confirmation which are independent of the neutral hydrogen fraction are desirable.

The CIII] and CIV rest-frame ultra-violet (UV) lines have been proposed as such alternatives \citep[e.g.][]{2010ApJ...719.1168E,2014MNRAS.445.3200S}.
Photoionization models predict that CIII] is the strongest rest-frame UV emission line at $\lambda<2700$\AA\ after \lya\ \citep{2016arXiv161003778J} with a ratio between CIV and CIII] often below one \citep{2016MNRAS.462.1757G,2016MNRAS.456.3354F}.
Therefore, even though the observed equivalent width (EW) ratios between CIII] and \lya\ are often very small \citep{2015ApJ...814L...6R}, models imply that CIII] is the most promising alternative to \lya\ for redshift determination in the early Universe.
In the near future, the James Webb Space Telescope (JWST) will enable independent redshift determinations from rest-frame optical emission lines at $\lambda\gtrsim2700$\AA\ of such sources, making non-\lya\ redshifts in the early universe more accessible.
However, rest-frame UV emission lines still provide valuable measures of physical conditions in star forming galaxies and Active Galactic Nuclei (AGN), and will likely continue to do so after the launch of JWST.

From photoionization models it is evident that the detection of rest-frame UV lines (other than \lya) is not only useful for redshift confirmation, but also for determining the physical properties of the emitting galaxies \citep[e.g.,][]{2010ApJ...719.1168E,2016ApJ...826..159S,2017MNRAS.tmp..190P,2017NatAs...1E..52A}.
For instance, as the rest-frame UV emission lines are closely related to the electron density, the gas-phase metallicity and the ionization state of the underlying stellar populations of the galaxies, they can provide constraints which are complimentary to rest-frame optical spectra.
By determining fluxes and flux ratios (or the corresponding limits) in the rest-frame UV, studies of galaxy properties in individual galaxies at the EoR have recently started to emerge \citep{2015MNRAS.454.1393S,2015MNRAS.450.1846S,2017MNRAS.464..469S,2017ApJ...836L..14M} painting a picture of highly ionizing and star forming systems with low metallicity.    
Hence, the continued search for rest-frame UV emission lines has the potential to not only provide confirmation of the redshifts of EoR systems when \lya\ is not available, but will also be useful to help determine the characteristics of the galaxies that reionized the IGM at $z>6$.
Rest-frame UV emission lines will therefore be invaluable beacons of information when assembling the picture of the very early Universe.

Here we present the detection of CIV$\lambda\lambda$1548,1551\AA\ from \HST\ slitless spectroscopy from the Grism Lens-Amplified Survey from Space \citep[GLASS\footnote{\url{http://glass.astro.ucla.edu/}};][]{2014ApJ...782L..36S,2015ApJ...812..114T} in a multiply imaged LAE at the EoR.
\lya\ and CIV detections from this object were also presented in 
\cite{2013A&A...559L...1B,2013A&A...559L...9B,2015A&A...574A..11K,2016ApJ...818...38S} and \cite{2017ApJ...836L..14M}.
We use the GLASS measurement, together with spectroscopic limits on the CIII]$\lambda\lambda$1907,1909\AA\ and HeII$\lambda$1640\AA\ lines, to explore the physical properties of the galaxy via recent photoionization models.
In Section~\ref{sec:sys} we present the system and summarize the spectroscopic confirmations in the literature. 
In Section~\ref{sec:data} we describe the \HST\ spectroscopy and photometry as well as the lens models used in this work. This leads to a discussion of whether the proposed images of the lensed system indeed belong to the same background object in Section~\ref{sec:images}. We confirm that one of the six proposed images (Image F) is likely not part of the system.
In Section~\ref{sec:UVEL} we present the CIV detections as well as upper limits and marginal detections for other UV emission lines from the combined and individual components of the multiply imaged system.
In Section~\ref{sec:mainali} we compare our result with the recent study by \cite{2017ApJ...836L..14M} who presented an independent CIV detection in one component of the system before we conclude our study in Section~\ref{sec:conc}.
Throughout this paper magnitudes are given in the AB magnitude system by \cite{1983ApJ...266..713O} and a standard cosmology with $H_0 = 70$km/s/Mpc, $\Omega_m = 0.3$, and $\Omega_\Lambda = 0.7$. 

\section{A Multiply Imaged LAE at $\lowercase{z}=6.11$}
\label{sec:sys}

We study the multiply imaged system behind the galaxy cluster \RXJ\ \citep[also known as Abell S1063;][]{1989ApJS...70....1A} at $z=0.348$ \citep{2004A&A...425..367B} first presented by \cite{2014MNRAS.438.1417M}.
From modeling the spectral energy distribution (SED) using \cite{2003MNRAS.344.1000B} models with a \cite{2003PASP..115..763C} initial mass function and Padova 1994 evolutionary tracks, \cite{2014MNRAS.438.1417M} estimated the background source to be a low-metalicity ($\textrm{Z}<0.2\textrm{Z}_\odot$) $\sim10^8 M_\odot$ star-bursting galaxy at $z\sim5.9$.
Several components of this system have since been spectroscopically confirmed via detection of the \lya\ line at $z=6.11$.
\cite{2013A&A...559L...1B} and \cite{2013A&A...559L...9B} were the first to spectroscopically confirm components of the system.
These detections were corroborated by \cite{2015A&A...574A..11K}, \cite{2016ApJ...818...38S} and \cite{2017ApJ...836L..14M}.
This makes the background LAE one of very few multiply imaged systems at the EoR with multiple images spectroscopically confirmed.
At present, only 3 other spectroscopically confirmed systems are known at $z>6$ \citep{2011MNRAS.414L..31R,2014ApJ...783L..12V,2016arXiv161201526V,2016ApJ...823L..14H}.

\begin{figure*}
\begin{center}
\includegraphics[width=0.99\textwidth]{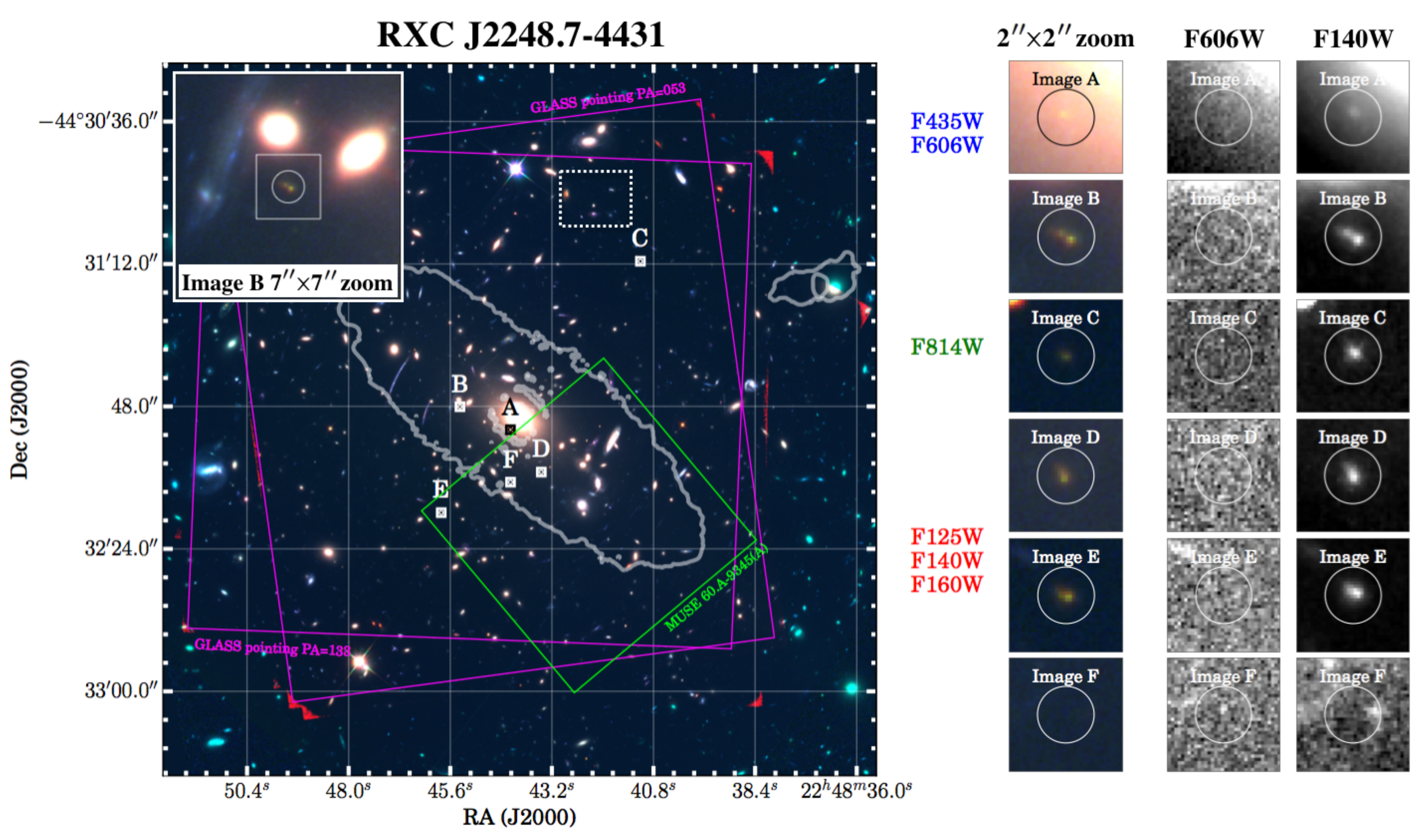}
\caption{False-color of image \RXJ\ with the components of the multiply imaged system at redshift 6.11 marked.
The left panel shows the \RXJ\ cluster core where the blue, green and red channels contain the HFF images listed to the right of the panel.
The insert in the upper left corner zooms in on Image B, showing the two nearest cluster members removed when estimating the photometry for Image B.
The field-of-view of the two GLASS pointings and the MUSE pointing described in the text are marked by the magenta and green squares, respectively.
The white contours show the critical curve at $z=6.11$ from our SWUnited lens model.
The small boxes show the location and size of the $2\farcs0\times2\farcs0$ ($\sim 11\textrm{kpc}\times11 \textrm{kpc}$ at $z=6.11$) postage stamps shown on the right.
The false-color postage stamps are comprised of the same HFF data as the full field-of-view image on the left.  
The grayscale postage stamps show the F606W epoch 1 (center) and F140W epoch 2 (right) HFF data.
We note that given the location of Images A-E, the location of image F is at odds with the lensing configuration predicted by the lens model 
\citep[as also pointed out by][]{2015A&A...574A..11K}, confirming
that image F is likely not belonging to the multiply lensed system of the $z=6.11$ object seen in the other images. 
The dashed rectangle to the north-east of image~C marks the location of where a counter image to image~F would be located according to the HFF lens models, should image~F depict another LAE at $z\sim6$.
See Section~\ref{sec:imgF} for the full discussion on image~F.
}
\label{fig:systemimg}
\end{center}
\end{figure*}

\cite{2014MNRAS.438.1417M} initially presented five images of the background LAE.
Later, \cite{2015A&A...574A..11K} suggested a tentative sixth image of the system (named Image F, here) from an emission line detection in Multi Unit Spectroscopic Explorer (MUSE) data.
\cite{2015A&A...574A..11K} emphasized that the location of this image is in disagreement with their lens model's predictions. We will reexamine this conclusion in  Section~\ref{sec:imgF}.
In Table~\ref{tab:objID} we provide the coordinates of all components of the system proposed in the literature and give references to the IDs used in the different studies.
The position of the individual images are marked on the false-color image of \RXJ\ in Figure~\ref{fig:systemimg}, where we also show postage stamps zoomed in on each image.

\tabletypesize{\small} \tabcolsep=0.11cm
\begin{deluxetable*}{lllllccccccccc} \tablecolumns{13}
\tablewidth{0pt} \tablecaption{Literature IDs of The $z=6.11$ Multiply Imaged System Behind \RXJ}
\tablehead{\colhead{Image} & 
\colhead{RA$_\textrm{J2000}$} &
\colhead{Dec$_\textrm{J2000}$} &
\colhead{$\alpha_\textrm{J2000}$} &
\colhead{$\delta_\textrm{J2000}$} &
\colhead{ID} &  
\colhead{ID} &  
\colhead{ID} &  
\colhead{ID} &  
\colhead{ID} &  
\colhead{ID} &  
\colhead{ID} &  
\colhead{ID} &  
\colhead{ID} \\  
\colhead{} & 
\colhead{[deg]} &
\colhead{[deg]} &
\colhead{[hms]} &
\colhead{[dms]} &
\colhead{Sc16} &
\colhead{CLASH} &
\colhead{Ba13} &
\colhead{Bo13} &
\colhead{Mo14} &
\colhead{Ri14} &
\colhead{Jo14} &
\colhead{Ka15} &
\colhead{Ma17}
}
\startdata
\hline \\[-1.5ex]
A      &    342.18408     & -44.5316378   & 22:48:44.179  &  -44:31:53.90   &   \nodata      &  \nodata   &      ID5       &    ID0      &     ID0         &  \nodata     & \nodata    & \nodata      & \nodata       \\
B      &    342.1890479 &  -44.5300249  & 22:48:45.37    & -44:31:48.18    &    01131       &   00847    &       ID1      &     ID1     &      ID1        &    6.1          &  12.1        &  \nodata     &  \nodata       \\
C      &    342.171296   &  -44.519812    & 22:48:41.11    &  -44:31:11.32    &    01752       &   00401    &       ID4      &     ID4     &      ID4        &    6.4          &  12.4        &  \nodata     &  \nodata       \\
D      &    342.18104     & -44.53461       & 22:48:43.45    &  -44:32:04.63   &    00845       &   01154    &       ID2      &     ID2      &     ID2        &     6.3         &   12.2       &   53a          &   \nodata     \\
E     &    342.190889   & -44.537461     & 22:48:45.81    &  -44:32:14.89   &    00699        &  01291     &     ID3       &   ID3        &   ID3           &    6.2          &  12.3        &  53b           &  ID3              \\
F$^*$ & 342.18407    &  -44.53532      & 22:48:44.177  &  -44:32:07.14   &     \nodata    &    \nodata  &     \nodata  &   \nodata &     \nodata &     \nodata  &  \nodata  &    53c         &    \nodata        
\enddata   
\tablecomments{The coordinates and different IDs used in the literature for the components of the multiply imaged object behind \RXJ. 
The IDs refer to
Sc16: \cite{2016ApJ...818...38S}; 
CLASH: The NIR catalogs release by the CLASH collaboration at \url{https://archive.stsci.edu/prepds/clash/};
Ba13: \cite{2013A&A...559L...9B}; 
Bo13: \cite{2013A&A...559L...1B};  
Mo14: \cite{2014MNRAS.438.1417M}; 
Ri14: \cite{2014MNRAS.444..268R}; 
Jo14: \cite{2014ApJ...797...48J}; 
Ka15: \cite{2015A&A...574A..11K}; 
Ma17: \cite{2017ApJ...836L..14M}; 
$^*$As explained in Section~\ref{sec:imgF}, we verified that image F is most likely not part of the system.
}
\label{tab:objID}
\end{deluxetable*}

\section{\HST\ Spectroscopy, Photometry, SEDs and Lens Models}
\label{sec:data}

In this section we summarize the data used in the current study.
These include Hubble Space Telescope (\HST) grism spectroscopy, archival \HST\ broad band imaging as well as archival lens models based on existing photometry and spectroscopic redshifts.

\subsection{The Grism Lens-Amplified Survey from Space (GLASS)}
\label{sec:glass}

The main spectroscopy used in this work are from near-infrared (NIR) \HST\ slitless grism spectroscopy acquired as part of GLASS.
GLASS observed the cores of 10 massive galaxy clusters with the G102 and G141 Wide-Field Camera 3 (WFC3) grisms covering the wavelength range $\lambda = 0.8-1.7\mu$m.
Each cluster was observed at two distinct position angles (PAs) roughly 90 degrees apart (indicated by the magenta squares in Figure~\ref{fig:systemimg}) to accommodate robust contamination subtraction and confirmation of low signal-to-noise ratio (S/N) emission lines.
The GLASS spectroscopy is reaching 1$\sigma$ line flux sensitivities of background sources of roughly $5\times10^{-18}$erg/s/cm$^2$/$\mu$, where $\mu$ is the lens magnification from the foreground cluster at the position of the source.
The broad NIR wavelength range enables the detection of rest-frame UV emission lines, in particular \lya, of a large number of $z\gtrsim6$ galaxies as illustrated by the 24 LAEs at $z\gtrsim7$ presented by \cite{2016ApJ...818...38S}, the multiply imaged LAE at $z=6.85$ presented by \cite{2016ApJ...823L..14H} as well as the current study.
The GLASS data were reduced using an updated version of the 3D-HST reduction pipeline \citep{2016ApJS..225...27M}.
For further details on the GLASS survey and data reduction we refer the reader to \cite{2014ApJ...782L..36S} and \cite{2015ApJ...812..114T}.
Figure~\ref{fig:individualspec} in Appendix~\ref{app:individualspec} shows the 24 individual 2D spectra of the six proposed components of the $z=6.11$ system in G102 and G141 at the two distinct PAs.

\subsection{HFF Imaging and Updated Photometry}
\label{sec:phot}

The cluster \RXJ was initially shown to be a prominent lens by the The Cluster Lensing And Supernova survey with Hubble \citep[CLASH;][]{Postman:2012ca}. These data enabled the discovery and study of the $z=6.11$ LAE studied here as explained in Section~\ref{sec:sys},
 and qualified the cluster to be selected as one of the six massive efficient galaxy clusters studied in the Hubble Frontier Fields campaign \citep[HFF;][]{2016arXiv160506567L}.
As part of the HFF, \RXJ\ was observed multiple times with \HST\ from late 2014 to mid 2016, obtaining deep broad-band images in the optical and NIR with limiting magnitudes down to m$_\textrm{AB}\sim28.6$ (not accounted for lensing magnification).
For further details about the HFF data\footnote{\url{http://www.stsci.edu/hst/campaigns/frontier-fields/}} we refer to \cite{2016arXiv160506567L}.
The completed full-depth HFF imaging was combined with ancillary imaging from the GLASS program, the CLASH imaging, and the \HST\ program ID 14209 (PI: B. Siana).
The \HST\ imaging was complemented by Spitzer IRAC channel 1--2 data from the Spitzer Frontier Fields campaign\footnote{\url{http://ssc.spitzer.caltech.edu/warmmission/scheduling/approvedprograms/ddt/frontier/}} (Capak et al. in prep.)
containing $\sim$50 hr per band in channel 1 [3.6] and channel 2 [4.5]. 
We did not include archival observations in channel 3 [5.7] and channel 4 [7.9] as these limits are too shallow to put meaningful constraints on the SEDs presented in Section~\ref{sec:SEDs}.
All these data result in two broad-band images blue-wards of the Lyman limit, three in-between the Lyman limit and the \lya\ wavelength, and nine broad-bands including or being red-wards of the \lya\ wavelength for this object (cf. Table~\ref{tab:obj}).
The two Spitzer bands cover wavelengths long-ward of the Balmer Break.

We estimated the photometry for each component of the multiply imaged system, from the combined mosaics of all these data,
many of which are made publicly available by the HFF team.\footnote{\url{https://archive.stsci.edu/prepds/frontier/abells1063.html}}
The photometric measurements were obtained by closely following the approach adopted and developed for the ASTRODEEP catalogs for A2744 and MACS0416 \citep{2016A&A...590A..30M,2016A&A...590A..31C}.
The ASTRODEEP pipeline models the intra-cluster light (ICL) and bright cluster galaxies with S\'ersic and Ferrer components using \verb+GALFIT+ \citep{2002AJ....124..266P} and subtracts these, before detecting objects and measuring fluxes, improving photometry of sources close to bright cluster members (e.g., image A and B of the $z=6.11$ system studied here, see Figure~\ref{fig:systemimg}).
We ran the source detection on the F160W mosaic and obtained broad band fluxes in dual-image mode using \verb+SExtractor+ \citep{1996A&AS..117..393B}.
As flux errors are often underestimated in deep co-added images due to correlated noise, we estimate the flux error of each source by placing empty apertures within 0$\farcs$5 to 2$\farcs$0 from each source, similar to the approach described by \cite{2011ApJ...727L..39T}.

For each image we compared three different estimates of the final magnitudes: 
\begin{itemize}
\item[i)] Nominal isophotal magnitudes.
\item[ii)] Nominal isophotal magnitudes imposing ACS magnitude limits for low-S/N detections.
\item[iii)] Magnitudes estimated in a 0$\farcs$4 diameter circular aperture with ACS magnitude limits imposed on the low-S/N detections.
\end{itemize}
All magnitudes are aperture corrected, where the aperture correction is defined as the difference between SExtractors estimate of the total magnitude (\verb+MAG_AUTO+) and the isophotal magnitude (\verb+MAG_ISO+) which is more appropriate for measureing colors.
In Table~\ref{tab:obj} we list the magnitudes obtained from ii) for images B-F, whereas the magnitudes for Image A are aperture magnitudes from iii). 
The photometry of image A is, despite the \verb+GALFIT+ cluster member subtraction and ICL modeling, affected by the bright central cluster galaxy \RXJ (see Section~\ref{sec:imgA}) biasing isophotal magnitudes.
We determined the validity of the aperture photometry and aperture corrections for image A based on fake-source simulations in the vicinity of image A.
We found iii) to be the most robust photometric measurement for image A and have therefore quoted these results in Table~\ref{tab:obj}.

\tabletypesize{\scriptsize} \tabcolsep=0.2cm
\begin{deluxetable*}{llccccccc} \tablecolumns{9}
\tablewidth{0pt} \tablecaption{Characteristics of Proposed Images of the Multiply Imaged $z=6.11$ LAE Behind \RXJ}
\tablehead{\colhead{} & 
\colhead{} &
\colhead{PA} &
\colhead{Image A} & 
\colhead{Image B} & 
\colhead{Image C} & 
\colhead{Image D} & 
\colhead{Image E} & 
\colhead{Image F} 
}
\startdata
\hline \\[-1.5ex]
$z_\textrm{spec}$      &   {\cite{2013A&A...559L...1B}} &	&   \nodata            &  Yes                     	&  Yes                   &  Yes        &  Yes       & \nodata                        \\
$z_\textrm{spec}$      &   {\cite{2013A&A...559L...9B}} & 	&  \nodata             &   \nodata	                 &  Yes                   &  Yes        &  Yes       &\nodata                         \\
$z_\textrm{spec}$      &  {\cite{2015A&A...574A..11K}} & 	&    \nodata           &  \nodata                 	 &  \nodata               &  Yes        &  Yes       &  Yes                             \\
$z_\textrm{spec}$      &   {\cite{2016ApJ...818...38S}}  &	&    \nodata           &  Yes                          &  Yes	               &  Yes        &  Yes        & \nodata                         \\
$z_\textrm{spec}$      &   {\cite{2017ApJ...836L..14M}} &  	&    \nodata           &  \nodata                      &  \nodata	     &  \nodata	  &  Yes       &  \nodata                \\       
[0.1cm]\hline \\[-1.5ex]
$\mu_\textrm{HFF}$    & ($\mu_\textrm{median}\pm75\%$ range) & &  1.46 $^{+12.09}_{-1.10}$  &  4.19 $^{+4.88}_{-2.79}$ & 1.95 $^{+0.36}_{-0.60}$ &   3.19 $^{+3.07}_{-1.82}$ &  4.22 $^{+0.76}_{-1.51}$ &  9.34 $^{+7.95}_{-3.32}$ \\
$\mu_\textrm{SWU}$  & ($\mu_\textrm{best}$ [68\% range]) & &  0.25[0.30,0.30] &  0.97[1.17,1.18] &  1.76[1.70,1.79] &  1.15[1.39,1.39]  &  3.14[2.60,2.63] &  3.55[6.35,6.77]  \\
[0.1cm]\hline \\[-1.5ex]
F435W 	& [AB mag]   &    &     $>$30.45 		&  $>$29.87 		&  $>$29.95  		& $>$30.15  		& $>$29.78 		&  $>$30.64  \\
F475W 	& [AB mag]   &    &     $>$28.49 		&  $>$28.28 		&  $>$28.35  		& $>$26.51  		& $>$28.26 		&  $>$28.75  \\
F606W 	& [AB mag]   &    &     $>$30.01 		&  $>$29.47 		&  $>$30.02  		& $>$29.40  		& $>$29.55 		&  $>$29.93  \\
F625W 	& [AB mag]   &    &     $>$28.35 		&  $>$27.28		&  $>$28.50  		& $>$26.99  		& $>$27.89 		&  $>$28.18  \\
F775W 	& [AB mag]   &    &     $>$27.60 		&  $>$27.62 		&  $>$28.01  		& $>$26.16  		& $>$27.88 		&  $>$28.16  \\
F814W 	& [AB mag]   &    &     27.07 $\pm$0.23  	&  26.18 $\pm$0.08  	& 26.94 $\pm$0.05  	&  26.23 $\pm$0.04  	& 26.04 $\pm$0.03  	&  $>$29.06  \\
F850LP 	& [AB mag]   &    &    26.75 $\pm$0.42  	&  25.36 $\pm$0.12  	& 25.76 $\pm$0.14  	&  24.81 $\pm$0.11  	& 25.06 $\pm$0.09  	&  $>$27.49  \\
F105W 	& [AB mag]   &    &     26.18 $\pm$0.09  	&  25.04 $\pm$0.04  	& 25.84 $\pm$0.02  	&  25.16 $\pm$0.02  	& 24.93 $\pm$0.01  	&  28.03 $\pm$0.20  \\
F110W 	& [AB mag]   &    &    26.26 $\pm$0.23  	&  24.88 $\pm$0.09  	& 26.45 $\pm$0.14  	&  25.21 $\pm$0.05  	& 25.00 $\pm$0.03  	&  $>$28.44  \\
F125W 	& [AB mag]   &    &     26.05 $\pm$0.17 	&  25.05 $\pm$0.05 	& 25.83 $\pm$0.04  	&  25.18 $\pm$0.02  	& 24.93 $\pm$0.02  	&  28.03 $\pm$0.22  \\
F140W 	& [AB mag]   &    &     26.12 $\pm$0.24  	&  25.14 $\pm$0.07  	& 25.95 $\pm$0.04  	&  25.28 $\pm$0.03  	& 25.07 $\pm$0.02  	&  28.19 $\pm$0.23  \\
F160W 	& [AB mag]   &    &     25.95 $\pm$0.13  	&  25.17 $\pm$0.09  	& 25.98 $\pm$0.04  	&  25.30 $\pm$0.03  	& 25.10 $\pm$0.02  	&  28.06 $\pm$0.25  \\
$[3.6]$	& [AB mag]   &    &    $>$24.66 			&  $>$22.34		&  24.73$\pm$0.28	&  $>$25.40		&  23.78$\pm$0.10	&  $>$25.97 \\
$[4.5]$	& [AB mag]   &    &    $>$24.64 			&  $>$22.65		&  25.24$\pm$0.41 	&  $>$25.52		&  24.76$\pm$0.25	&  $>$25.99 \\
[0.1cm]\hline \\[-1.5ex]
$M_\textrm{1500}$ 	& [AB mag]   &    &     $-20.13 ^{+8.28}_{-0.76}$  	&  $-20.12 ^{+1.17}_{-0.67}$  	& -$20.15 ^{+0.19}_{-0.31}$  	&  $-20.30^{+0.96}_{-0.57}$  	& $-20.23^{+0.18}_{-0.36}$  	&  $-16.26 ^{+0.88}_{-0.43}$ \\
[0.1cm]\hline \\[-1.5ex]
 $f_\textrm{Ly$\alpha$}$   &   [1e-17 erg/s/cm$^2$] &  053  &  1.93 $\pm$ 1.32 &  $<$ 1.30 & 3.02 $\pm$ 0.98 & 5.31 $\pm$ 1.04 & 5.14 $\pm$ 1.01 &  $<$ 1.35  \\
 $f_\textrm{Ly$\alpha$}$   &   [1e-17 erg/s/cm$^2$] &  133  &  2.08 $\pm$ 1.50 &  $<$ 1.46 & 1.83 $\pm$ 0.91 & 5.58 $\pm$ 0.96 & 5.49 $\pm$ 0.95 & 3.04 $\pm$ 1.05  \\
 $f_\textrm{NV}$   &   [1e-17 erg/s/cm$^2$] &  053  &   $<$ 1.13 &  $<$ 1.09 & 2.74 $\pm$ 0.80 &  $<$ 1.09 & 2.29 $\pm$ 0.81 & 1.82 $\pm$ 0.85  \\
 $f_\textrm{NV}$   &   [1e-17 erg/s/cm$^2$] &  133  &   $<$ 1.15 &  $<$ 1.18 &  $<$ 0.93 &  $<$ 0.97 & 1.27 $\pm$ 0.96 &  $<$ 0.93  \\
 $f_\textrm{CIV}$   &   [1e-17 erg/s/cm$^2$] &  053  &  0.66 $\pm$ 0.32 &  $<$ 0.54 & 0.59 $\pm$ 0.48 & 1.56 $\pm$ 0.42 & 2.59 $\pm$ 0.48 &  $<$ 0.49  \\
 $f_\textrm{CIV}$   &   [1e-17 erg/s/cm$^2$] &  133  &   $<$ 0.51 &  $<$ 0.44 &  $<$ 0.43 &  $<$ 0.94 & 0.83 $\pm$ 0.37 & 0.67 $\pm$ 0.43  \\
 $f_\textrm{HeII}$   &   [1e-17 erg/s/cm$^2$] &  053  &   $<$ 0.99 &  $<$ 1.04 &  $<$ 0.70 &  $<$ 0.98 &  $<$ 0.63 &  $<$ 0.95  \\
 $f_\textrm{HeII}$   &   [1e-17 erg/s/cm$^2$] &  133  &   $<$ 1.05 &  $<$ 1.12 &  $<$ 0.77 &  $<$ 0.77 &  $<$ 0.79 &  $<$ 0.57  \\
 $f_\textrm{OIII]}$   &   [1e-17 erg/s/cm$^2$] &  053  &   $<$ 0.98 &  $<$ 0.96 &  $<$ 0.86 &  $<$ 0.86 &  $<$ 0.85 &  $<$ 0.87  \\
 $f_\textrm{OIII]}$   &   [1e-17 erg/s/cm$^2$] &  133  &   $<$ 0.97 &  $<$ 1.00 &  $<$ 0.69 &  $<$ 0.77 &  $<$ 0.70 &  $<$ 0.72  \\
 $f_\textrm{CIII]}$   &   [1e-17 erg/s/cm$^2$] &  053  &   $<$ 0.76 &  $<$ 0.70 &  $<$ 0.58 &  $<$ 0.61 &  $<$ 0.58 &  $<$ 0.58  \\
 $f_\textrm{CIII]}$   &   [1e-17 erg/s/cm$^2$] &  133  &   $<$ 0.68 &  $<$ 0.39 &  $<$ 0.35 &  $<$ 0.51 &  $<$ 0.49 &  $<$ 0.50  \\
[0.1cm]\hline \\[-1.5ex]
 EW$_\textrm{Ly$\alpha$}$  &   [\AA]                &  053  &  83 $\pm$ 57 &  $<$ 19 & 94 $\pm$ 31 & 88 $\pm$ 17 & 69 $\pm$ 14 &  $<$ 315 \\
 EW$_\textrm{Ly$\alpha$}$  &   [\AA]                &  133  &  89 $\pm$ 64 &  $<$ 22 & 57 $\pm$ 28 & 93 $\pm$ 16 & 74 $\pm$ 13 & 711 $\pm$ 278 \\
 EW$_\textrm{NV}$  &   [\AA]                &  053  &   $<$ 48 &  $<$ 16 & 85 $\pm$ 25 &  $<$ 18 & 31 $\pm$ 11 & 425 $\pm$ 213 \\
 EW$_\textrm{NV}$  &   [\AA]                &  133  &   $<$ 49 &  $<$ 18 &  $<$ 29 &  $<$ 16 & 17 $\pm$ 13 &  $<$ 217 \\
 EW$_\textrm{CIV}$  &   [\AA]                &  053  &  35 $\pm$ 18 &  $<$ 11 & 25 $\pm$ 21 & 37 $\pm$ 10 & 49 $\pm$ 9 &  $<$ 161 \\
 EW$_\textrm{CIV}$  &   [\AA]                &  133  &   $<$ 27 &  $<$ 9 &  $<$ 19 &  $<$ 22 & 16 $\pm$ 7 & 221 $\pm$ 148 \\
 EW$_\textrm{HeII}$  &   [\AA]                &  053  &   $<$ 69 &  $<$ 30 &  $<$ 42 &  $<$ 32 &  $<$ 17 &  $<$ 450 \\
 EW$_\textrm{HeII}$  &   [\AA]                &  133  &   $<$ 74 &  $<$ 32 &  $<$ 46 &  $<$ 25 &  $<$ 21 &  $<$ 270 \\
 EW$_\textrm{OIII]}$  &   [\AA]                &  053  &   $<$ 69 &  $<$ 27 &  $<$ 51 &  $<$ 28 &  $<$ 23 &  $<$ 412 \\
 EW$_\textrm{OIII]}$  &   [\AA]                &  133  &   $<$ 68 &  $<$ 28 &  $<$ 41 &  $<$ 25 &  $<$ 19 &  $<$ 339 \\
 EW$_\textrm{CIII]}$  &   [\AA]                &  053  &   $<$ 56 &  $<$ 25 &  $<$ 44 &  $<$ 24 &  $<$ 19 &  $<$ 298 \\
 EW$_\textrm{CIII]}$  &   [\AA]                &  133  &   $<$ 50 &  $<$ 14 &  $<$ 26 &  $<$ 20 &  $<$ 16 &  $<$ 254 
\enddata   
\tablecomments{Characteristics of the components of the multiply imaged object behind \RXJ.
The ``$z_\textrm{spec}$'' rows list the publications presenting spectroscopic redshifts for the different components of the system. 
The lens magnifications, $\mu_\textrm{HFF}$, were estimated using the HFF calculator at \url{http://archive.stsci.edu/prepds/frontier/lensmodels/webtool/magnif.html}. 
The values quoted are the median magnification from the available models of \RXJ: CATS v1, Sharon v2, Zitrin-NFW v1, Zitrin-LTM v1, Zitrin-LTM-Gauss v1, Williams v1, Bradac v1 (SWUnited), and Merten v1 (see HFF webtool for details). 
The error-bars represent the range of the model predictions after removing the largest and smallest magnification values, i.e., ignoring 2/8 models which roughly corresponds to 75\% confidence intervals.
The magnifications for our SWUnited model ($\mu_\textrm{SWU}$) and their 68\% confidence intervals correspond to the critical curve shown in Figure~\ref{fig:systemimg}.
Magnitudes are from updated ICL-subtracted photometry based on a combination of HFF, CLASH and GLASS imaging (see Section~\ref{sec:phot} for details).
All magnitude limits and uncertainties are 1$\sigma$.
$M_\textrm{1500}$ is estimated from the F105W magnitude ($M_{1500} = \textrm{F105W} - \textrm{DL}_\textrm{6.11} - 2.5 * \log10(1 + 6.11)$ where $\textrm{DL}_\textrm{6.11}$ is the distance modulus at $z=6.11$) and has been corrected for the lens magnification ($M_\textrm{1500} = M_\textrm{1500,obs.} + 2.5 * \log10(\mu_\textrm{HFF})$).
Emission line fluxes and rest-frame EWs are obtained from ellipsoidal aperture measurements on the individual 2D grism spectra from GLASS.
Also flux and EW limits and uncertainties are 1$\sigma$.
The combined fluxes and rest-frame EWs are given in Table~\ref{tab:obj_comb} and described in Section~\ref{sec:fcorrection}.
}
\label{tab:obj}
\end{deluxetable*}

Comparing our updated photometry from combining archival imaging with the HFF imaging described above, to the archival photometry from CLASH\footnote{\url{https://archive.stsci.edu/prepds/clash/}}, the magnitudes for images C, D, and E, which are less affected by light from neighboring cluster members, agree to within $\sim0.2$ mag averaged over the bands F850LP, F105W, F110W, F125W, F140W, and F160W, where all objects are detected.
The CLASH team also provided measurements for image B. 
In this case the updated ICL-subtracted magnitudes presented here differ from the previous values by on average $\sim$0.8 magnitudes.
This is to be expected as we subtracted two bright neighboring cluster members (see insert in Figure~\ref{fig:systemimg}) during the \verb+GALFIT+ modeling, which have likely contaminated the CLASH photometry of image B.
The CLASH catalogs do not contain measurements of images A and F due to contamination from the bright central galaxy of \RXJ\ and a non-detection in the shallower CLASH imaging, respectively.

The IRAC imaging used here was deep enough to detect image~C in band [3.6] and image~E in both [3.6] and [4.5]. 
The remaining images only have upper limits to their IRAC fluxes due to heavy blending with foreground galaxies or ICL in IRAC (images~A, B and D) or non-detections (images F).
Detections with IRAC have the power of probing the optical rest-frame emission.
If all the IRAC flux was produced by starlight the [4.5] flux is expected to be comparable to, or greater than the flux in [3.6].
An excess in the [3.6] band can be attributed to a contribution from strong
[OIII]$\lambda\lambda$4959,5007\AA\ line emission at $z=6.11$.  
Such excesses have been proven to be very useful for the selection and study of high-redshift star-forming galaxies \citep[e.g.,][Malkan et al. 2016, ApJ in press]{2013ApJ...763..129S,2014ApJ...784...58S,2015ApJ...801..122S,2016ApJ...823..143R}.
Both image~C and E show a strong flux excess in the [3.6] band. We will discuss this further in Section~\ref{sec:SEDs} below.

The UV slope, $\beta$, based on the obtained photometry for images A-E has a median value of $\beta=-2.32$. 
Combining the UV values for the individual images accounting for uncertainties in the estimates by perturbing the measured flux values according to their photometric errors we get $\beta\sim-2.37\pm0.04$ (68\% confidence intervals).
This is not quite as steep as the UV slope $\beta=-2.89\pm0.38$ estimated by \cite{2014MNRAS.438.1417M}, and is therefore in better agreement with measurements from HUDF09, ERS and CANDELS \citep{Bouwens:2012ht,2014ApJ...793..115B,2012ApJ...756..164F} correlating the absolute UV magnitude with the spectral slope, $\beta$. 
For $M_\textrm{UV}=-20.2$ (mean of magnification-corrected $M_\textrm{UV}$ for images A-F) at redshift 6 these relations predict a $\beta$-value of roughly -2.0, though with a significant scatter in the individually measured slopes of up to $\pm$1.

Using \verb+GALFIT+ we find a median half-light radius measured along the major axis of images B--E 
(uncorrected for lens-distortion) in the F160W images of $\sim$350pc.
Image A and F were too contaminated/faint to reveal sensible measurements.
This size measurements indicates that the background LAE is fairly compact in the continuum, and is in agreement with estimated sizes of equally bright galaxies at similar redshift from the literature \citep[e.g.,][]{2013ApJ...765...68H,2015ApJS..219...15S,2015ApJ...804..103K,2016ApJ...821L..27V,2016ApJ...825...41V,2016arXiv161201526V}.

\subsection{SED Modeling of Photometry}
\label{sec:SEDs}

\begin{figure*}
\begin{center}
\includegraphics[width=0.49\textwidth]{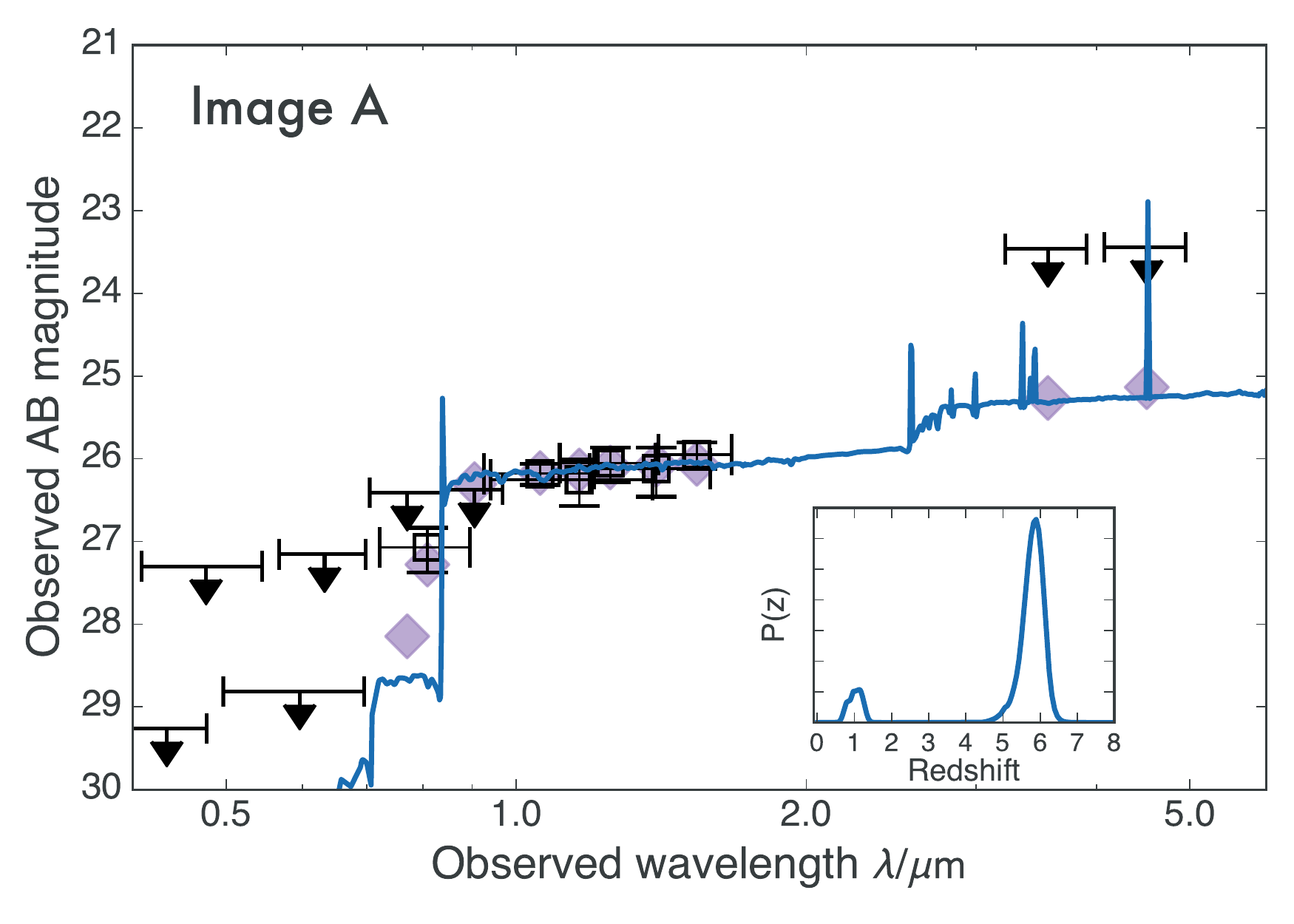}
\includegraphics[width=0.49\textwidth]{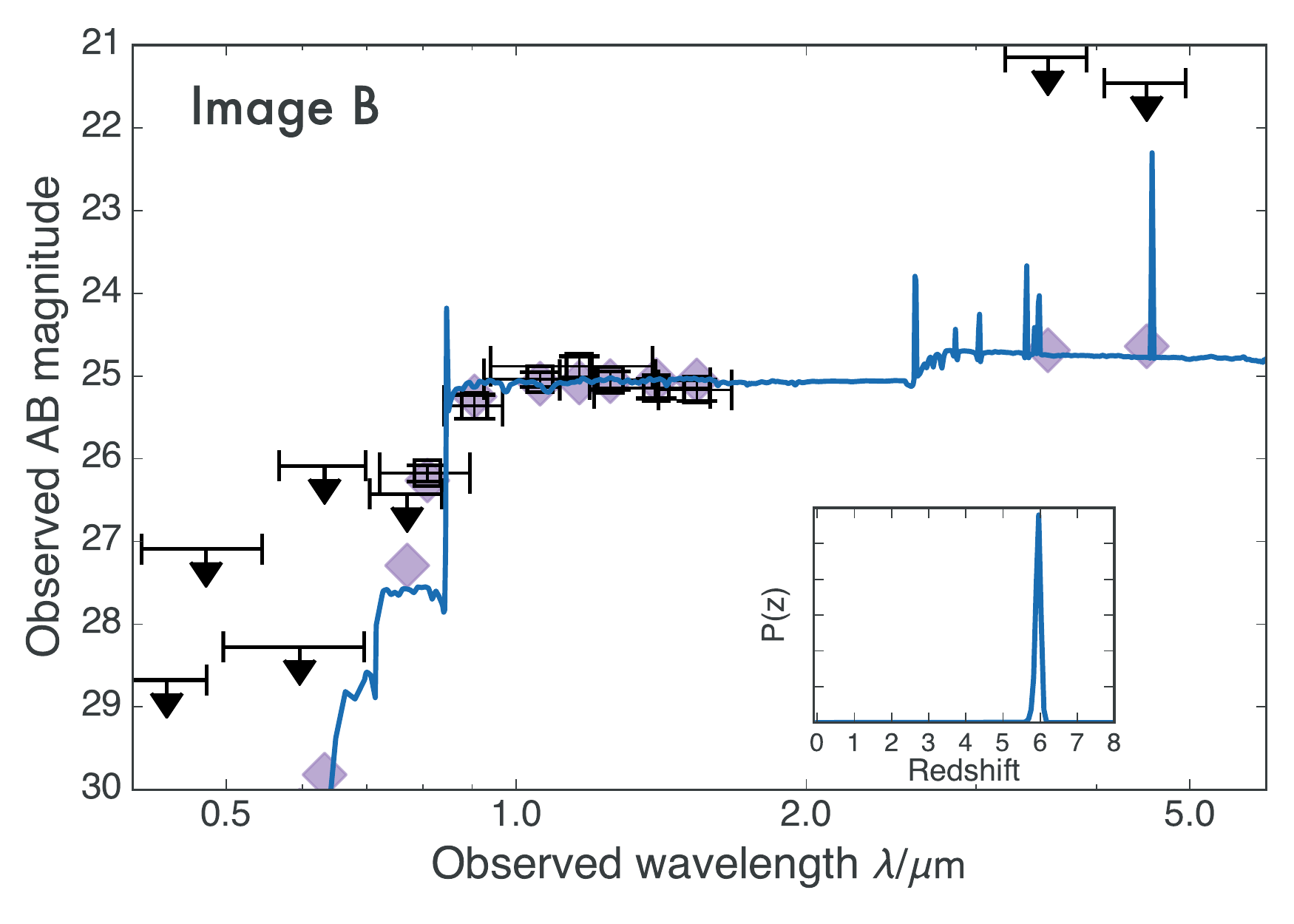}\\
\includegraphics[width=0.49\textwidth]{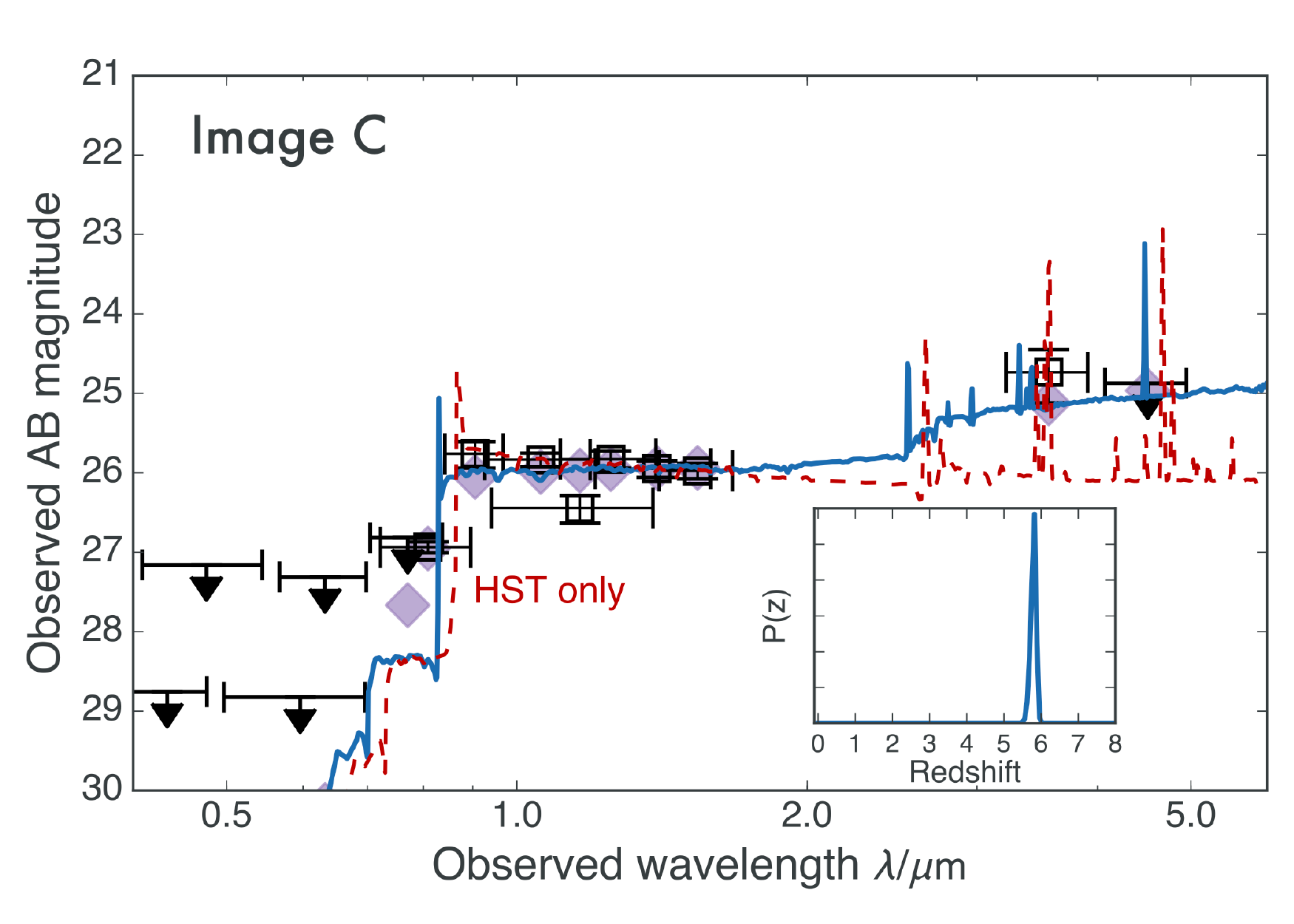}
\includegraphics[width=0.49\textwidth]{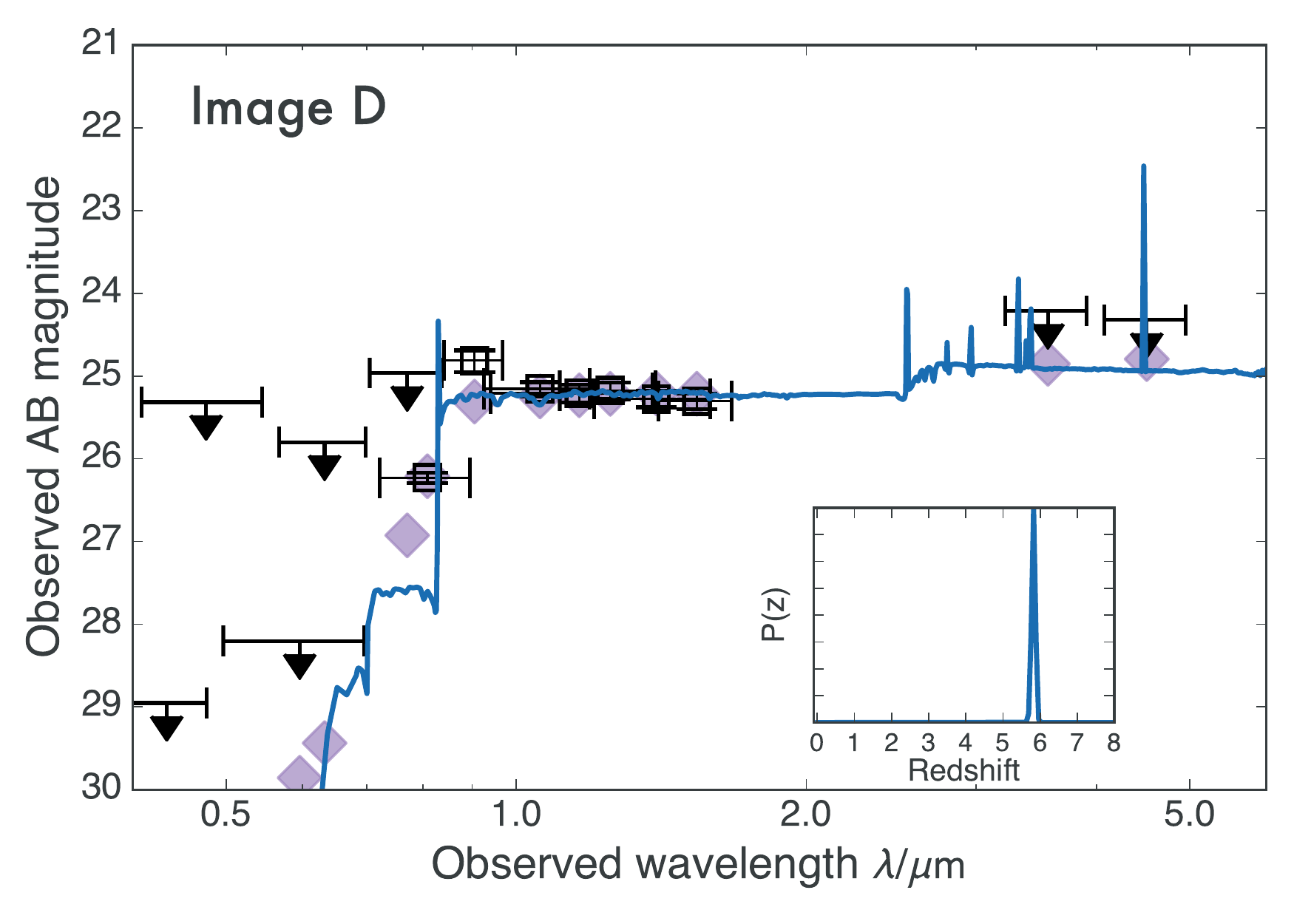}\\
\includegraphics[width=0.49\textwidth]{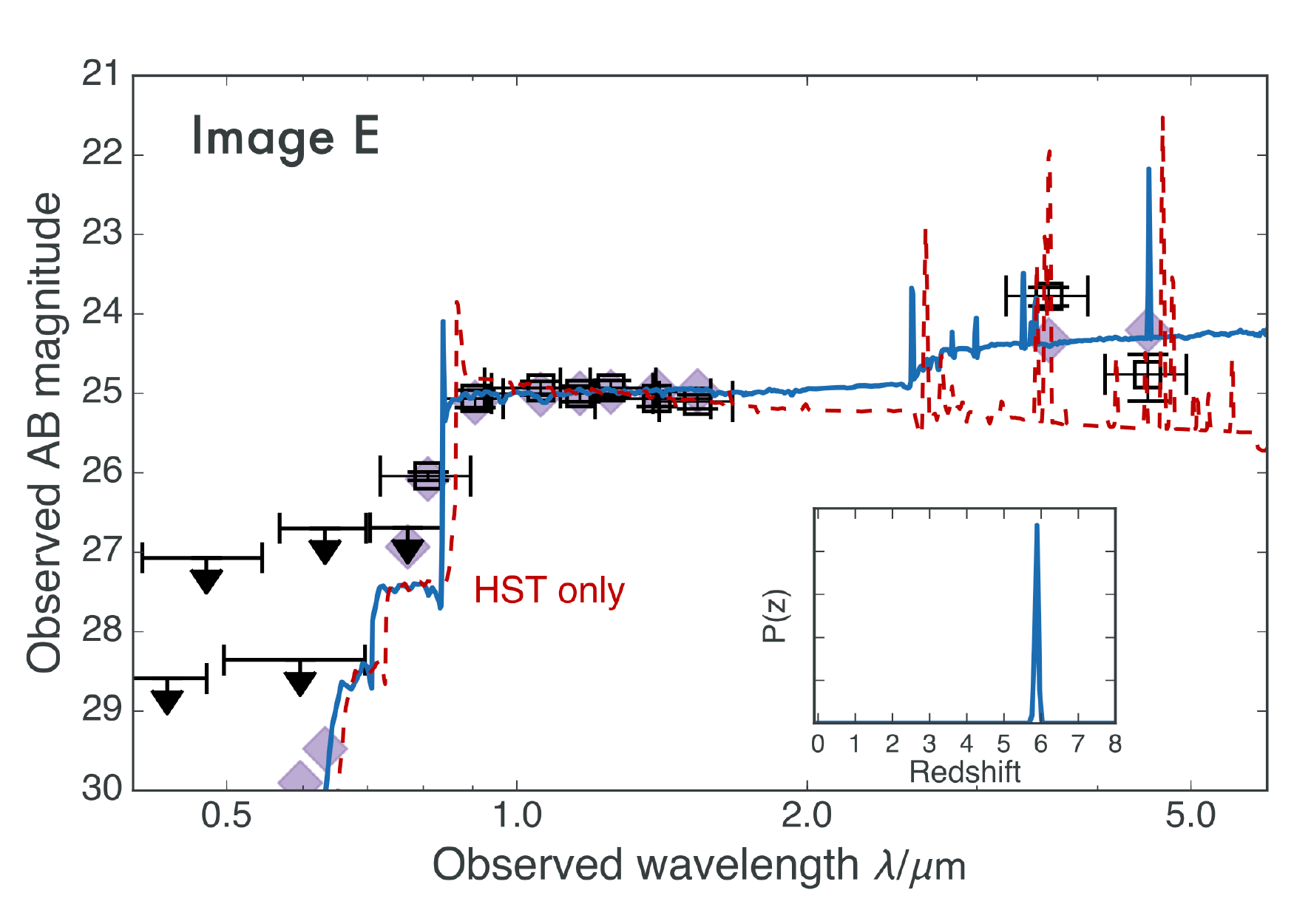}
\includegraphics[width=0.49\textwidth]{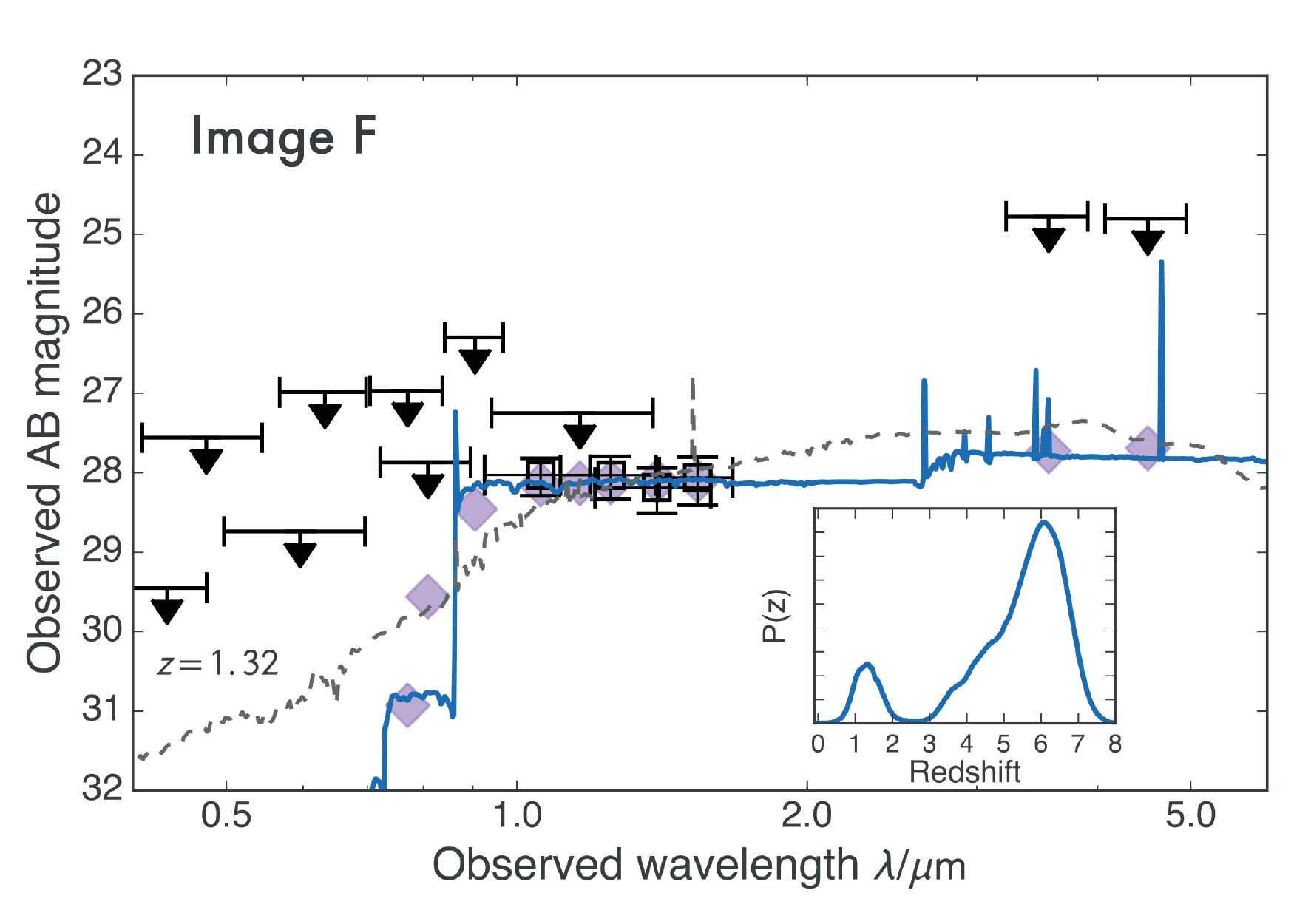}
\caption{
The photometric measurements (black symbols) and the corresponding best-fit SEDs (blue curves) of images A-F.
Black squares and black arrows show the magnitudes and magnitude 3$\sigma$ upper limits for each of the photometric bands listed in Table~\ref{tab:obj}.
The purple diamonds show the magnitudes predicted by the best-fit SED.
Images B-E all have narrow $p(z)$ distributions (inserted panels) around a redshift solution of $z\sim6$ in agreement with the spectroscopic confirmation of these images.
Image A and F have bimodal probability distributions indicating non-negligible probabilities that the photometry shows a $z\sim1$ object.
Fixing the redshift at $z=6.11$ we obtain stellar mass, star formation rate, reddening and ages for all six images as shown in Figure~\ref{fig:physparam}.
The red dashed SEDs in the panels for Image~C and E (center and bottom left), show the best-fit SEDs with a young stellar population fit to \HST\ data only, where the IRAC fluxes and colors are accounting for with strong nebular rest-frame optical emission lines.
Including IRAC photometry, these solutions formally have a worse fit than SEDs with old stellar populations for images Image~C and E, but are potentially a better solution considering the look-back time of only $\sim$900Myr at $z=6.11$.
See Section~\ref{sec:SEDs} for details.
In the panel for Image~F (bottom right), the black dashed SED shows the best-fitting solution with $z=1.32$.
At $z\sim1.32$ the [OII]$\lambda\lambda$3726,3729\AA\ doublet falls at the wavelength of \lya\ at $z=6.11$.
As described in Section~\ref{sec:imgF} the potential detection in the GLASS spectra and in the available MUSE data, might be a detection of [OII] rather than \lya, in agreement with the SED fit show here.
}
\label{fig:SEDs}
\end{center}
\end{figure*}

Using the updated ICL-subtracted photometry we estimated the photometric redshift of each of the components of the multiply imaged system using EA$z$Y \citep{2008ApJ...686.1503B} with SED models from \cite{2003MNRAS.344.1000B},
a \cite{2003PASP..115..763C} initial mass function,
and the dust attenuation curve from \cite{2000ApJ...533..682C}.
The resulting best-fit SEDs are shown for images A-F in Figure~\ref{fig:SEDs}. 
The redshift probability distribution, $p(z)$, for each fit is shown in the inserted panels. 
All images favor an SED solution at $z\sim6$, in agreement with the spectroscopic redshifts of the individual images presented here and in the literature.
Image A and F are the only images with a non-negligible probability of being a galaxy at lower redshift according to these SED fits. 
We will discuss each of the SED fits in Section~\ref{sec:images} and investigate the low-redshift solution for image F further in Section~\ref{sec:imgF}.

\begin{figure*}
\begin{center}
\includegraphics[width=0.19\textwidth]{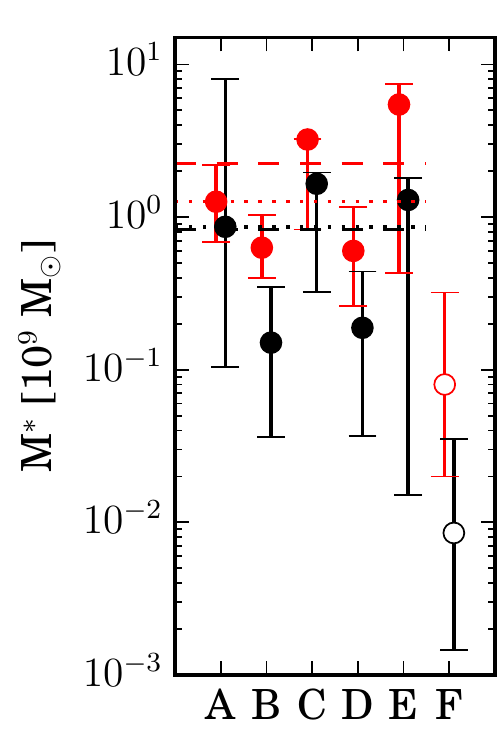}
\includegraphics[width=0.19\textwidth]{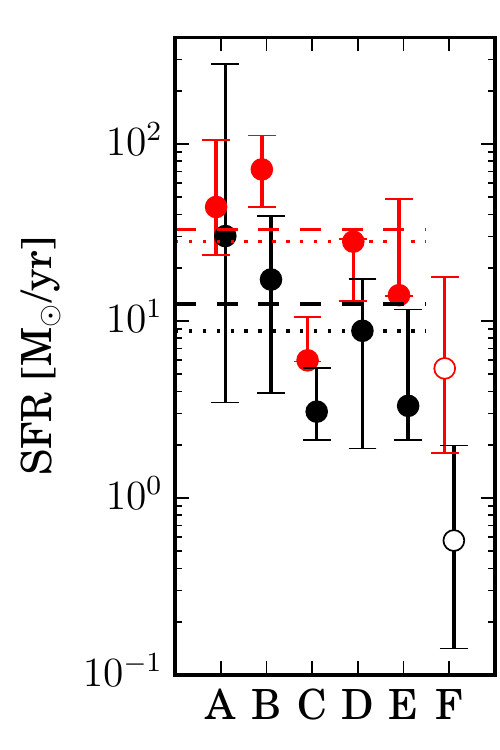}
\includegraphics[width=0.19\textwidth]{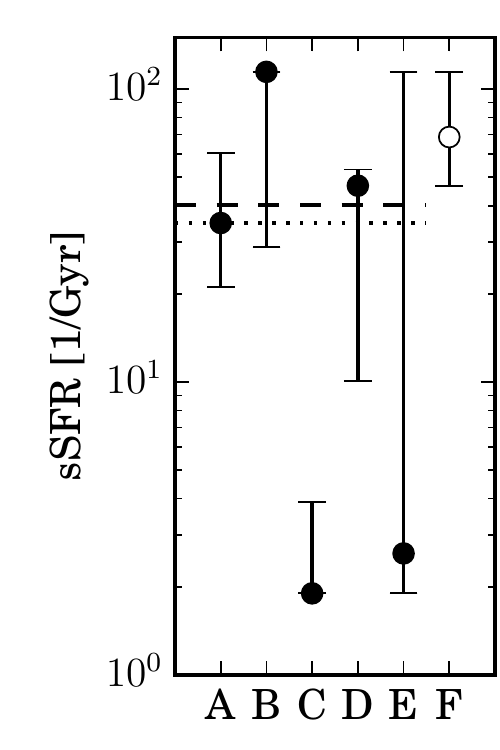}
\includegraphics[width=0.19\textwidth]{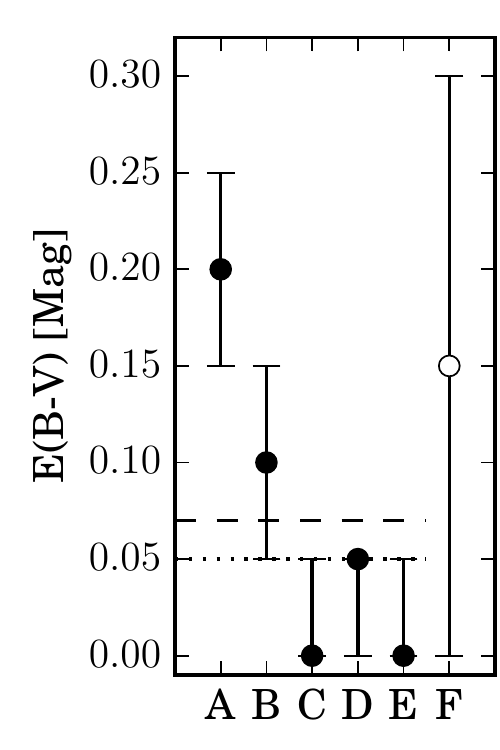}
\includegraphics[width=0.19\textwidth]{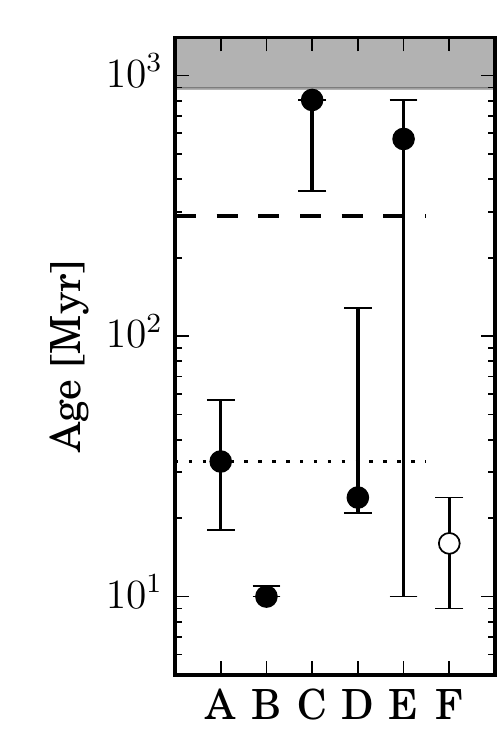}
\caption{
The physical parameters obtained from SED template fitting to the photometry shown in Table~\ref{tab:obj} fixing the redshift to $z=6.11$ for the Images A-F (x-axis).
M$^*$ and SFR have been corrected for the lensing magnification $\mu_\textrm{HFF}$.
The uncorrected values are shown by the red symbols (applying a horizontal shift to the points for clarity). 
The sSFR, E(B-V) and age are independent of the lens magnification and these panels therefore only contain black symbols.
The error bars on each of the estimates show the 68\% ($\sim$1$\sigma$) confidence intervals added the uncertainty on the lens magnification in quadrature when necessary (M$^*$ and SFR).
The horizontal dashed (dotted) lines mark the mean (median) values for images A-E.
The gray shaded band in the Age panel marks unphysical population ages larger than the look-back time at $z=6.11$.
The measurements for image F (which formally agree with the medians from images A--E in several cases, but are very uncertain) are marked with open circles, 
as we confirm that this objects is likely not part of the multiply imaged system
as detailed in Section~\ref{sec:imgF}.
The M$^*$ and SFR agrees for images A-E within a few $\sigma$.
The same is true for the sSFR, except for image C which has an estimated sSFR below 3.9Gyr$^{-1}$ (1$\sigma$) driven by the low SFR from this SED fit.
The low E(B-V) values preferred by the SED fits are in agreement with previously inferred values in the literature and are all within 2$\sigma$ of the average value.
The best-fit ages for image~C and E are large. This is due to the fact that very high EW rest-frame optical emission lines are needed to fit the IRAC colors with a young stellar population as discussed in Section~\ref{sec:SEDs}. However, apart from image C, the preferred age is below 50Myr.
}
\label{fig:physparam}
\end{center}
\end{figure*}

Fixing the redshift of the SED fits to $z=6.11$ we used the \cite{2003MNRAS.344.1000B} models assuming a constant star formation history and 0.2 solar metallicity \citep[the upper bound reported by][]{2014MNRAS.438.1417M} we get estimates for the stellar mass (M$^*$), the star formation rate (SFR), the specific star formation rate (sSFR), the reddening (E(B$-$V)), and the age of the underlying stellar population (Age; the minimum age of the templates in the library is 10Myr).
The results from these fits are shown for all six images in Figure~\ref{fig:physparam}.
The M$^*$ and SFR are dependent on the lens magnification and have been corrected by $\mu_\textrm{HFF}$ described in Section~\ref{sec:lensmodel} (adding the uncertainties in quadrature).
The red symbols in Figure~\ref{fig:physparam} show the values from the observed photometry before correcting for the lens magnification, $\mu_\textrm{HFF}$.
We note that the results are essentially identical if we use isophotal magnitudes (i) instead of isophotal magnitudes with ACS magnitude limits (ii) for images B-F.

Overall the lens magnification corrected values of M$^*$ and SFR agree for images A-E within a few $\sigma$.
Compared to the stellar mass estimate presented by \cite{2014MNRAS.438.1417M} and \cite{2017ApJ...836L..14M}, the updated photometry presented here (ignoring image F) appears to prefer a higher mass of the lensed source ($\sim10^9$M$_\odot$ as opposed to $\sim10^8$M$_\odot$).
The SFR is predicted to be of the order 10M$_\odot$/yr.
The sSFR shown in Figure~\ref{fig:physparam} agree within 1$\sigma$, except for the estimated sSFR for image C which is below 3.9Gyr$^{-1}$ (1$\sigma$), whereas the mean and median values are closer to 35Gyr$^{-1}$ (1$\sigma$).
This is in good agreement with the 50Gyr$^{-1}$ presented for Image~E by \cite{2017ApJ...836L..14M}.
The model for the evolution of the galaxy UV luminosity function by \cite{2015ApJ...813...21M} predicts a sSFR of 3Gyr$^{-1}$ for a galaxy of M$^*\sim10^9$M$_\odot$ at $z\sim6$ with an assumed stellar mass to halo mass ratio of 0.01.
This is somewhat below the best-fit sSFR for the $z\sim6.11$ objects, but is in fair agreement considering the uncertainties in both the models and the SED fits.
The low values of the reddening, E(B-V), preferred by the SED fits also agree within a few $\sigma$ and are in agreement with the values found by \cite{2014MNRAS.438.1417M} (A$_\textrm{V} \sim 0.2-0.4$) and the E(B-V)=0.01 found for Image~E by \cite{2017ApJ...836L..14M}.

The main difference in the fitted SED parameters appear in the estimated age of the underlying stellar population. 
Images A, B and D, for which we are only able to measure IRAC upper limits, seem to prefer a stellar population younger than $\lesssim 30$Myr. 
Images C and E also favor stellar population templates younger than 50Myr old when only \HST\ flux densities are included in the SED fits (these \HST-only fits are shown by the red dashed lines in Figure~\ref{fig:SEDs} for images C and E).
However, when their IRAC detections are included, both are favored by older ($> 500$Myr old) templates, even though they are both still consistent with very young ($< 50$Myr old) templates within $1\sigma$. 
At $z=6.11$, both [OIII]$\lambda\lambda$4959,5007\AA\ and H$\beta$ fall in the IRAC [3.6] bandpass, and H$\alpha$ falls in the [4.5] band. The blue [3.6]$-$[4.5] colors of images C and E ($-0.51 \pm 0.5$ and $-0.98 \pm 0.27$, respectively) suggest strong nebular emission lines \citep[e.g.,][]{2015ApJ...801..122S,2016ApJ...817...11H} and very young stellar populations. 
But our \cite{2003MNRAS.344.1000B} templates have a hard time matching the red F160W$-$[3.6] and F160W$-$[4.5] colors even with nebular emission lines of large EW$_\textrm{[OIII]} > $1000 and EW$_\textrm{H$\alpha$} >$ 500 and line ratios from \cite{2003A&A...401.1063A} included. 

There are two possible reasons why our templates can not simultaneously reproduce the blue [3.6]$-$[4.5] color and the red F160W$-$IRAC colors. The first possibility is that this galaxy has emission line ratio [OIII]/H$\alpha$ much higher than those inferred by \cite{2003A&A...401.1063A}, in agreement with the expected progressively increasing [OIII]/H$\alpha$ ratio at $3<z<6$ \citep{2016ApJ...821..122F}.
The second possibility is that the Lyman-continuum photon production rate predicted by the \cite{2003MNRAS.344.1000B} models are too low for this galaxy. 
Recent studies \citep{2016ApJ...831..176B} argue for a high LyC photon production rate per UV luminosity at $z \sim 4-5$. This could also explain the red F160W$-$IRAC colors for a stellar population younger than 50Myr.
Otherwise the red F160W$-$IRAC colors suggest at least a moderate 4000\AA break.

To summarize, the SED fits to the photometry of images A--E of the $z=6.11$ LAE indicate a galaxy with a mass of a few times 10$^9$M$_\odot$, a SFR of 50--100M$_\odot$/yr, low dust content, and a preferred stellar population $<50$Myr old.
\subsection{Lens Models of \RXJ}
\label{sec:lensmodel}

Being part of the HFF efforts and having extensive spectroscopy from GLASS\footnote{\url{https://archive.stsci.edu/prepds/glass/}}, 
CLASH-VLT \citep{2013A&A...559L...9B,2016A&A...587A..80C} and 
MUSE \citep{2015A&A...574A..11K}, several high-fidelity lens models of \RXJ\ exist \citep[e.g.,][and \url{https://archive.stsci.edu/prepds/frontier/abells1063_models_display.html}]{2014MNRAS.444..268R,2014ApJ...797...48J,2016A&A...587A..80C}.
In Figure~\ref{fig:systemimg} we show the critical curve at $z=6.11$ (white contour) from our Strong and Weak Lensing United \citep[{SWUnited},][]{2009ApJ...706.1201B,2005A&A...437...39B} lens model, which is based on spectroscopic redshifts from \cite{2014MNRAS.444..268R} and \cite{2014MNRAS.438.1417M}, i.e. pre-HFF imaging and pre-GLASS spectroscopy.

To estimate the lensing magnification at the location of each of the images A--F we use the HFF magnification calculator\footnote{\url{https://archive.stsci.edu/prepds/frontier/lensmodels/\#magcalc}} which returns the predictions from all the available HFF lens models.
As large scatter in the estimated magnifications between different models can occur, in Table~\ref{tab:obj} we quote the median magnification at each image position ($\mu_\textrm{HFF}$) after removing the minimum and maximum value.
The error bars represent the range of magnifications in the remaining models which roughly corresponds to a 75\% confidence interval ($\sim$$1\sigma$) given that two of the eight estimates were ignored. 
We also quote the values from our SWUnited model explicitly ($\mu_\textrm{SWU}$).
We will use the combined HFF estimates as the lens magnification for each image in the remainder of this paper unless mentioned otherwise.

\section{Is the $\lowercase{z}=6.11$ LAE Sextuply Imaged?}
\label{sec:images}

In light of the results presented in this paper, the updated HFF photometry on \RXJ\ and results presented in the recent literature, in this section, we will assess whether or not each of the proposed components is likely to be images of the same LAE at $z=6.11$.

\subsection{Images B--E}
\label{sec:imgBE}

Images B, C, D, and E were all independently confirmed to have \lya\ emission from ground-based spectroscopy at high enough resolution and S/N to resolve the line profile (cf. references in Table~\ref{tab:obj}).
Hence, there exists little doubt, if any, that these are all components of the same multiply imaged \lya\ emitter at $z=6.11$.
The best-fitting SEDs to the updated ICL-subtracted photometry and the corresponding narrow $p(z)$ profiles shown in Figure~\ref{fig:SEDs}, as well as the detection of \lya\ and CIV in the GLASS spectra from several of these components, which we will describe in Section~\ref{sec:UVEL}, confirm this conclusion.

\subsection{Image A}
\label{sec:imgA}

The SED fit to the updated ICL-subtracted photometry of image A shown in Figure~\ref{fig:SEDs} suggests a high redshift solution for this image, in agreement with the brighter, spectroscopically confirmed components of the system.
However, the $p(z)$ does suggest a non-negligible low-redshift solution for the SED-fit.
Due to image A's vicinity to the bright central galaxy of \RXJ\ we rely on aperture photometry to get as robust a magnitude measurement as possible.
The aperture photometry of image A quoted in Table~\ref{tab:obj} is in good agreement with the visual impression that image A is fainter than image B in the NIR (see Figure~\ref{fig:systemimg}).

The potential \lya\ detections in the spectra of image A are both below 2$\sigma$ and the single CIV detection is also tentative, but agree well with the preferred high-$z$ solution of the SED fits

Combined with the predictions of an image of the background LAE at the location of Image A by the lens models (we searched the vicinity of image A for other high-$z$ candidates but found none), these measurements leads us to believe that image A is likely a counterpart of the $z=6.11$ system as suggested in the literature.

The estimates of M$^*$, SFR, sSFR, E(B$-$V), and the age from the SED fitting shown in Figure~\ref{fig:physparam} are in good agreement with the average values of the other images (the mean and median values shown in Figure~\ref{fig:physparam} changes very little by leaving out Image A), and therefore seem to represent the same (intrinsic) photometry.

\subsection{Image F}
\label{sec:imgF}

\cite{2015A&A...574A..11K} proposed a 6th image of the $z=6.11$ multiply lensed system based on a tentative emission line detection at a wavelength agreeing with the wavelength of \lya\ in images D and E from the MUSE data of program ID 60.A-9345(A).
\cite{2015A&A...574A..11K} emphasized that the component and the detected emission has to be confirmed and that their lens model did not predict an image of the $z=6.11$ system at the observed location.
The critical curve at $z=6.11$ from our {SWUnited} lens model shown in Figure~\ref{fig:systemimg} confirms that image F is not expected to be part of the $z=6.11$ system, as the existence of an additional image at the location of image~F is incompatible with the geometry of the critical curve and the presence of images D and E, which are both unambiguously part of the multiple image system.
The location of the critical curves for the remaining available HFF lens models also disagree with Image~F being another image of the system.
Hence, based on the lens model it is more likely that the emission is from a low-redshift object, or from another LAE at $z\sim6$ behind \RXJ.
In the latter case we would expect another set of multiple images from this source. Based on the HFF lens models we locate a $18\farcs0\times14\farcs0$ box (dashed white box to the NE of image C in Figure~\ref{fig:systemimg}) in the WFC3/IR field-of-view where 5/7 models predict the counter-image of image F to be (which deviates from the position of images A-E).
Five sources located within this box have a photometric redshift estimate $z>5$ in our photometric catalog.
None of the GLASS spectra extracted at the location of these sources show signs of an emission line at $\lambda\sim8645$\AA.
This is not surprising given that all candidate counterparts have observed magnitudes $m_\textrm{F105W}>28.75$.
In a single source a flux excess at $\lambda\sim8400$\AA\ is present in one of the two G102 spectra and is not accounted for by the contamination model. We consider this feature unlikely to be an emission line from the $\textrm{F105W}=29.61\pm0.28$ source.

The SED fit shown in the bottom right panel of Figure~\ref{fig:SEDs} prefers a high-$z$ solution, but the probability distribution is much broader for image F than for images B--E and also permits a low-$z$ solution at $z=1$--2.
If the MUSE line emission presented by \cite{2015A&A...574A..11K} is [OII]$\lambda\lambda$3726,3729\AA\, this would correspond to a galaxy at $z\sim1.32$ matching this low-$z$ solution.
We note that the spectral slope of Image~F, if it is at $z=6.11$, estimated from the photometry is $-2.24^{+0.73}_{-0.88}$ in agreement with the slope of images A-E quoted in Section~\ref{sec:phot}, which is expected giving the similarity of the SEDs of image~F and those of images A-E.

\begin{figure}
\begin{center}
\includegraphics[width=0.49\textwidth]{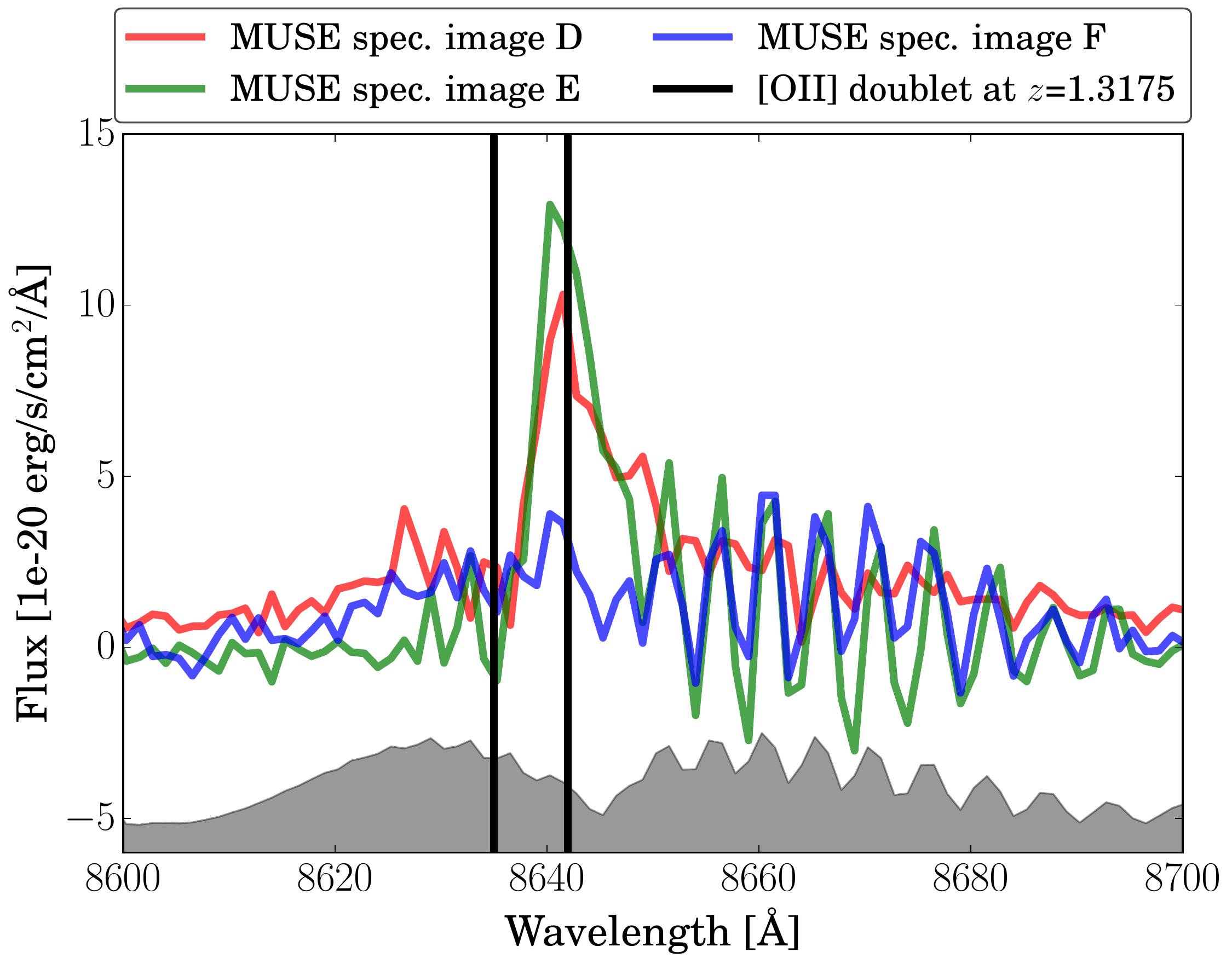}
\caption{
MUSE 1D spectra extracted from the publicly available data cubes from program ID 60.A-9345(A) of Images D, E and F. 
The spectra were extracted with a circular aperture with a size reflecting the seeing for the individual observations.
The sky level (offset and arbitrarily scaled) is shown by the gray shaded region.
From the line profiles in the MUSE spectra, it is clear that the detections at 8645\AA\ are \lya\ emission in images D and E (red and green profile, respectively). The potential marginal detection in the spectrum from image F (blue) might be [OII] emission at $z\sim1.3175$ marked by the vertical black lines and not \lya\ at $z\sim6.11$ as suggested by \cite{2015A&A...574A..11K}.
}
\label{fig:MUSEspec}
\end{center}
\end{figure}

As an independent check we downloaded the fully reduced MUSE data cubes of program ID 60.A-9345(A) available from the ESO phase 3 archive\footnote{\url{http://archive.eso.org/wdb/wdb/adp/phase3_main/form}} and extracted the 1D spectra of images D, E and F covered by these data.
The location of the MUSE field-of-view is shown by the green square in Figure~\ref{fig:systemimg}.  
We extracted the spectra using a simple circular aperture extraction setting the diameter of the extraction aperture to the seeing FWHM for each of the four individual exposures.
Figure~\ref{fig:MUSEspec} shows the resulting 1D spectra around the $z=6.11$ \lya\ wavelength. Images D and E show clear \lya\ profiles, whereas the detection in Image F is tentative. The profile of the emission does not match the expected skewed profile of \lya\ seen in images D and E.
This can be caused by a combination of the lower S/N, the increased sky emission in the vicinity of the line (shown in gray in Figure~\ref{fig:MUSEspec}), an underlying intrinsic non-\lya\ profile, or by the absence of a line altogether.
In Figure~\ref{fig:MUSEspec} we have shown the expected location of the [OII] doublet at $z=1.3175$ matching the low-redshift solution suggested by the bimodal $p(z)$ of the best-fit SED for image F shown in Figure~\ref{fig:SEDs}.
We do not detect any flux-excess at the location of the MgII$\lambda\lambda$2796,2803\AA\ doublet in the MUSE spectrum at this redshift, nor do we detect any [OIII]$\lambda\lambda$4959,5007\AA\ (which coincides with the location of HeII at $z=6.11$) or H$\alpha\lambda$6563\AA\ emission in the GLASS G141 spectra of Image~F.

As described in Section~\ref{sec:UVEL} below (cf. Table~\ref{tab:obj}) a flux excess at the $z=6.11$ \lya\ wavelength was detected in one of the G102 spectra of Image~F. This might be a confirmation of the low-S/N line (potential [OII] doublet) seen in the MUSE data.

At the proposed location of image F, the HFF lens models predict a higher magnification than at the location of images A-D (cf. Table~\ref{tab:obj}). This can be seen in Figure~\ref{fig:systemimg} by the proximity of image F to the critical curve at $z=6.11$.
Hence, if image F belongs to the same $z=6.11$ background source, it should appear brighter in both photometry and in the \lya\ line than the other less-magnified components of the system. 
This is clearly illustrated by the inferred absolute UV magnitude ($M_{1500}$) of Image~F listed in Table~\ref{tab:obj} which, given the estimated magnification and the faint observed magnitudes, is several magnitudes below $M_{1500}$ for images A-E.
Even though the individual lens models are uncertain, the data disagree with image F being the brightest image and it is therefore unlikely that image F is part of the multiply imaged system and that the feature seen in the MUSE and GLASS spectra is \lya\ of the same intrinsic strength observed in the other components.
The potential CIV detection (1.6$\sigma$) at PA=133 of Image F listed in Table~\ref{tab:obj} is tentative, and does therefore not contradict this conclusion.

Based on the above arguments, we do \emph{not} believe that image F is a multiple image of the $z\sim6.11$ LAE emitter behind \RXJ\ studied here, and we therefore excluded the spectra and measurements of image~F from any of the stacked or combined measurements presented in the analysis and discussion in the remainder of this paper.
This conclusion agrees with the caution that image F is not likely part of the system by \cite{2015A&A...574A..11K}.
Instead image F might be a low-redshift object with [OII] being emitted at the observed wavelength of \lya\ at $z=6.11$.

\vspace{0.5cm}
To conclude, according to the above, the $z=6.11$ LAE is lensed by \RXJ\ into a quintet of images A, B, C, D and E in the image plane of the cluster as originally identified by \cite{2014MNRAS.438.1417M} \citep[see also][]{2015A&A...574A..11K}.

\section{UV Emission Lines in the System}
\label{sec:UVEL}

To measure the line fluxes, flux limits and rest-frame EWs of the rest-frame UV lines in the GLASS spectra, we follow the approach detailed by \cite{2016ApJ...818...38S}. 
In short, we use ellipsoidal apertures applied to the contamination subtracted 2D spectra.
The sizes of these apertures are either manually optimized to maximize S/N for the flux measurements or set to a fixed size for the flux limit estimates.
The continuum levels used for the EW calculations are obtained from the HFF photometry described in Section~\ref{sec:phot}.
The line fluxes, 1$\sigma$ flux limits at the expected line locations, and the EW estimates are summarized in Table~\ref{tab:obj} for all components of the lensed system. Neither the flux measurements nor the flux limit estimates in Table~\ref{tab:obj} have been corrected for the lensing magnification, $\mu_\textrm{HFF}$.

\subsection{Combining Flux Measurements and Correcting for Lensing Magnification}
\label{sec:fcorrection}

The multiple imaging of the background source, and the multiple PAs of the GLASS observations, result in 10 independent spectra of the same source.
As listed in Table~\ref{tab:obj} the \lya\ and CIV emission lines were detected in multiple components of the system.
The lines were observed at wavelengths $\sim$8640\AA\ \citep{2016ApJ...818...38S} and $\sim$11010\AA.
Hence, the GLASS data does not indicate any offset between \lya\ and CIV in excess of the $\sim$24\AA\ dispersion of the G102 grism which corresponds to a velocity shift of $\sim830$km/s at the observed \lya\ wavelength. 

To combine the data to improve both S/N and reliability of the individual flux (limit) measurements we have taken two approaches:
i) we stacked the observed 2D spectra and ii) combined the individual measurements via a `probabilistic Gaussian product'. 
We will describe each of these approaches and the conclusions we can draw from them in the following.

\begin{figure*}
\begin{center}
\includegraphics[width=0.98\textwidth]{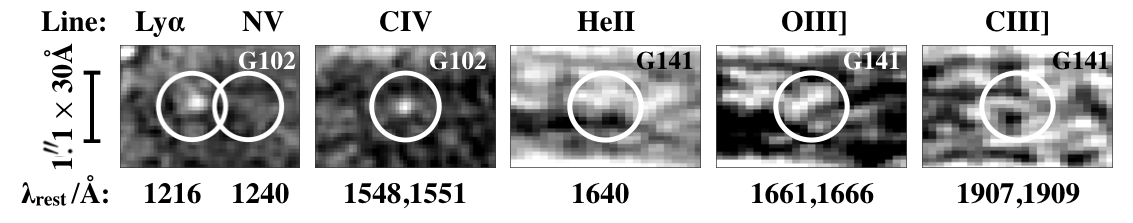}
\caption{
Cutouts of the rest-frame stacked GLASS spectra around the rest-frame UV emission lines \lya, NV, CIV, HeII, OIII] and CIII] of the $z=6.11$ LAE behind \RXJ.
The circles marking the position of the lines have diameters of 30\AA\ (rest-frame) and $1\farcs1$ in the dispersion (horizontal) and spatial (vertical) direction, respectively. 
Image B at PA=053 was excluded from the stack due to its severe contamination and significant contamination residual.
Image F was also excluded from the stack as this component is most likely not part of the system (cf. Section~\ref{sec:imgF}). 
The \lya\ and CIV lines are clearly detected in the stacks. 
The tentative detection of OIII] agrees with the detection presented by \cite{2017ApJ...836L..14M}.
}
\label{fig:stack}
\end{center}
\end{figure*}

In the first approach, we created rest-frame stacks of the contamination corrected 2D grism spectra to improve the \emph{observed} S/N of the emission lines.
Figure~\ref{fig:stack} shows 80\AA\ by 2$\farcs$0 cutouts from the inverse variance weighted stack around some of the expected rest-frame UV lines of the $z=6.11$ LAE.
The location of each line is marked by a white circle, which has a diameter of 30\AA\ (rest-frame) in the dispersion direction and $1\farcs1$ in the spatial direction.
Clear detections of \lya\ and CIV are seen in the rest-frame stacks. 
In an attempt to improve S/N and to remove any possible bias towards the brighter components of the multiply imaged system, we also generated: median stacks, stacks after weighting each spectrum by the measured \lya\ flux, stacks without weighting and/or scaling, stacks accounting for the continuum levels predicted by the \HST\ broad bands, and all possible combinations of these setups.
In all resulting stacks CIV (and \lya) is clearly detected.
In Figure~\ref{fig:stack} a low-S/N flux excess is seen at the OIII] doublet location, which is in agreement with the detection presented by \citep[][see Section~\ref{sec:mainali}]{2017ApJ...836L..14M}.
This feature and a potential flux excess at the location of CIII] show up in a few of the differently combined stacks, but deeper data \citep[like those presented by][]{2017ApJ...836L..14M} are needed to confirm these.

We measured the observed line fluxes and flux limits on these 2D rest-frame stacks. 
The measurements are summarized in Table~\ref{tab:obj_comb}.
Formally we detect OIII]$\lambda\lambda$1661,1666\AA\ at $\sim$3$\sigma$.
However, given significant contamination subtraction residual at the location of the OIII-doublet in the stacks, and given the lack of a detection in several of the alternative stacks described above (unlike \lya\ and CIV which were both clearly detected in all stacks) we consider this detection tentative.

\tabletypesize{\scriptsize} \tabcolsep=0.2cm
\begin{deluxetable}{lclc} \tablecolumns{9}
\tablewidth{0pt} \tablecaption{Combined Estimates of Line Flux and EW of the $z=6.11$ LAE Behind \RXJ\ (Excluding Image F)}
\tablehead{\colhead{} & 
\colhead{Unit} &
\colhead{Note} &
\colhead{Estimate} 
}
\startdata
\hline \\[-1.5ex]
$f_\textrm{Ly$\alpha$}$   	&   [1e-17 erg/s/cm$^2$] &  $\mu$-corrected   &  0.89$^{+0.24}_{-0.24}$ \\
$f_\textrm{NV}$$^\star$	&   [1e-17 erg/s/cm$^2$] &  $\mu$-corrected   &  0.43$^{+0.13}_{-0.13}$  \\
$f_\textrm{CIV}$   		&   [1e-17 erg/s/cm$^2$] &  $\mu$-corrected   &  0.21$^{+0.08}_{-0.04}$    \\
$f_\textrm{HeII}$   		&   [1e-17 erg/s/cm$^2$] &  $\mu$-corrected   &  $<0.21$   \\ 
$f_\textrm{OIII]}$   		&   [1e-17 erg/s/cm$^2$] &  $\mu$-corrected   &  $<0.25$   \\ 
$f_\textrm{CIII]}$   		&   [1e-17 erg/s/cm$^2$] &  $\mu$-corrected   &  $<0.15$   \\ 
[0.1cm]\hline \\[-1.5ex]
EW$_\textrm{Ly$\alpha$}$   	&   [\AA]  &  rest-frame & 68$^{+6}_{-6}$  \\
EW$_\textrm{NV}$$^\star$	&   [\AA]  & rest-frame  & 25$^{+5}_{-5}$  \\
EW$_\textrm{CIV}$   		&   [\AA]  & rest-frame  & 24$^{+4}_{-4}$  \\
EW$_\textrm{HeII}$   		&   [\AA]  & rest-frame  & $<26$  \\ 
EW$_\textrm{OIII]}$   		&   [\AA]  & rest-frame  & $<27$  \\ 
EW$_\textrm{CIII]}$   		&   [\AA]  & rest-frame  & $<20$ \\  
[0.1cm]\hline \\[-1.5ex]
$f_\textrm{Ly$\alpha$}$   &   [1e-17 erg/s/cm$^2$] &  stack (incl. $\mu$)  &  $1.95\pm0.39$ \\ 
$f_\textrm{NV}$   &   [1e-17 erg/s/cm$^2$] &  stack (incl. $\mu$)  &  $<0.46$\\
$f_\textrm{CIV}$   &   [1e-17 erg/s/cm$^2$] &  stack (incl. $\mu$)  & $0.89\pm0.15$\\
$f_\textrm{HeII}$   &   [1e-17 erg/s/cm$^2$] &  stack (incl. $\mu$)  & $<0.28$ \\
$f_\textrm{OIII]}$   &   [1e-17 erg/s/cm$^2$] &  stack (incl. $\mu$)  & $0.60\pm0.20$\\
$f_\textrm{CIII]}$   &   [1e-17 erg/s/cm$^2$] &  stack (incl. $\mu$)  & $<0.18$
\enddata   
\tablecomments{
Combination of fluxes and EWs from Table~\ref{tab:obj}.
The first set of fluxes were combined by taking the product of individual Gaussian probability density functions (PDF) representing both measurement and limits including correction for the estimated lensing magnification (see Section~\ref{sec:fcorrection} for details)
The combined EWs were estimated with the Gaussian PDF method.
The second set of fluxes at the bottom of the table were measured directly on the 2D stack shown in Figure~\ref{fig:stack}.
Scaling the \lya\ and CIV detections by the average HFF magnification ($\langle\mu_\textrm{HFF}\rangle=3.0$) the measurements on the stacks agree with the measurements from the Gaussian PDF combination.
Image F was excluded from the combined fluxes and EWs in all cases as this object is likely not part of the lensed system (cf. Section~\ref{sec:imgF}).
$^\star$As discussed in Section~\ref{sec:fEWinterp} the potential NV detection is very likely spurious.
}
\label{tab:obj_comb}
\end{deluxetable}

\begin{figure}
\begin{center}
\includegraphics[width=0.49\textwidth]{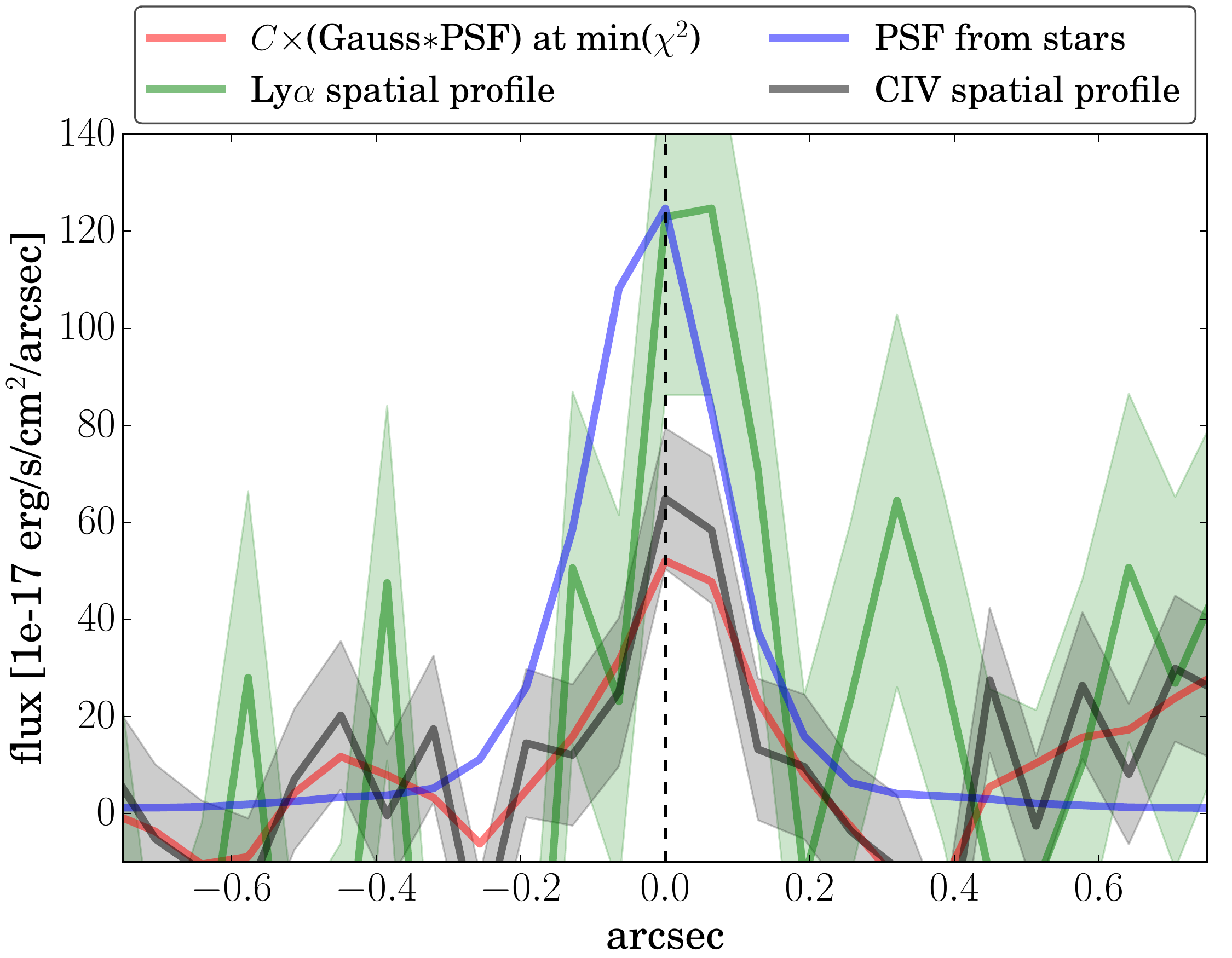}
\caption{
Comparison of the spatial profiles of \lya\ (green) and CIV (black) obtained from the stacked GLASS spectra (Figure~\ref{fig:stack}).
In blue the PSF obtained from a sample of stars in the GLASS fields is shown for comparison.
The red curve represents the Gaussian convolution of the CIV profile that minimizes the $\chi^2$ when compared to the \lya\ profile. This gives a quantitive measure of how different the profiles are \citep[see][for details]{2016ApJ...818...38S}.  
We do not detect any significant difference in the width of the two spatial profiles.
Both the CIV and \lya\ spatial profiles are un-resolved when compared to the PSF profile. 
Hence, deeper data are needed to draw any conclusion on the (dis)similarities of the extent of \lya\ and CIV emission lines which both resonantly scatter in the IGM.
}
\label{fig:spatprof}
\end{center}
\end{figure}

The rest-frame stack of the grism spectroscopy allows us to estimate the spatial extent of any detected emission in the 2D stacks.
Following the approach described by \cite{2016ApJ...818...38S}, 
we extracted spatial profiles of the \lya\ and CIV detections shown in the stacks in Figure~\ref{fig:stack}.
To represent the spatial profile of the PSF we extracted an average profile from GLASS spectra of bright stars.
Figure~\ref{fig:spatprof} shows the resulting spatial profiles of both the emission lines (green and gray curves) and the PSF (blue curve).
By modeling each of the spatial profiles as a Gaussian convolution of the PSF, sampling the standard deviations, and minimizing $\chi^2$ of the comparison we obtain a quantitative estimate of the difference between the emission line profiles themselves and the PSF \citep[red curve in Figure~\ref{fig:spatprof}; see][for details]{2016ApJ...818...38S}.
For both the \lya\ and CIV profile we do not find any significant deviation from the extent of the PSF. 
Hence, both emission line profiles are un-resolved in the GLASS observations and do not show any signs of an extended halo.
Compared to the surface brightness levels of known \lya\ halos beyond $0\farcs2$ \cite[$<10^{-17}$erg/s/cm$^2$/arcsec$^2$;][]{2016A&A...587A..98W} this is not surprising, as the GLASS stack is not expected to reach flux levels below $10^{-17}$erg/s/cm$^2$/arcsec$^2$.
Furthermore, as the morphology of the observed images are distorted by the foreground cluster lensing the stacked signal in the spatial direction is not necessarily enhanced by the same factor as the emission line (source) centers, further impeding the detection of the faint halos.

In the MUSE data described in Section~\ref{sec:imgF} the extent of the \lya\ flux is several times larger than the approximate $\textrm{FWHM}\sim0\farcs2$ reported in Figure~\ref{fig:spatprof}. 
This extent is driven mainly by the seeing of the MUSE observations ($1\farcs2$--$2\farcs2$ per visit), which makes it hard to directly compare the spatial extent to the one estimated here from the \HST\ data.
However, given that the MUSE data are reaching depths of $10^{-18}$erg/s/cm$^2$ \citep{2015A&A...574A..11K} a significant contribution from faint wings below the GLASS detection limits might be seen in the MUSE data.

\lya\ photons scatter in the IGM and circum-galactic medium, and extended \lya\ emission is therefore expected and has been reported on several occasions for both galaxies at $z>6$ as well as at lower redshifts \citep[e.g.,][]{Steidel:2011jk,Matsuda:2012fp,Momose:2014fe,2016A&A...587A..98W}.
How this compares to the CIV emission which also resonantly scatters \cite[e.g.][]{1983A&A...122..335F,1996A&A...312..751V,2007NewAR..51..194V,2013ApJS..207....1A,2015MNRAS.454.1393S} is still unknown, though we expect the CIV emission to be less extended, as HI probes (\lya) are more sensitive to low density gas, whereas metals (CIV) may be too weak to detect where overall gas column densities are low \citep{2015ApJ...809...19H,2016A&A...595A.100C}.
However, given that both the \lya\ and CIV emission is unresolved in the GLASS grism spectral stacks (and the MUSE wavelength range does not cover the CIV line), we can not draw any conclusions on the (dis)similarities of the two emitting regions.
Approved integral field unit observations with The K-band Multi-Object Spectrometer (KMOS) on VLT (196.A-0778, P.I. Fontana) will further help characterize the spatial extent of the CIV emission.

The second approach for combining the GLASS measurements, was to combine the individual fluxes ($f\pm\delta f$) and flux limits ($<\delta f$) listed in Table~\ref{tab:obj}.
We represented each measurement by a Gaussian probability density function (PDF) with mean $\langle f \rangle=f$ and standard deviation $\sigma=\delta f$. For the flux limits we used the 1$\sigma$ upper limits as $\langle f \rangle$ as well as $\sigma$ for the PDFs.
All PDFs are set to 0 for negative fluxes.
Each of these Gaussian PDF representations of the measurements are corrected by the lens magnifications listed in Table~\ref{tab:obj}, using that the intrinsic flux is simply the observed flux divided by the magnification, and that the variance on the intrinsic flux is given by:
\begin{equation}
\delta f^2_\textrm{intrinsic} = \frac{1}{\mu^2} \left( \frac{f^2_\textrm{observed} \; \delta \mu^2}{\mu^2} + \delta f^2_\textrm{observed} \right)
\end{equation}
Here $\mu\pm\delta\mu$ is the estimated lensing magnification for each image. The uncertainties on the magnification factors quoted in Table~\ref{tab:obj} are not symmetric, and we therefore, conservatively, use the largest absolute value of the upper/lower magnification ranges as $\delta\mu$.
The product of the individual magnification-corrected PDFs for each image gives us a representation of the intrinsic flux of the background source including information from both the detections and the upper limits. 
If any of the detections and/or upper flux limits were in disagreement such a product would be driven to 0. Within 3$\sigma$ all our measurements are in agreement with each other.
Sampling the distribution resulting from taking the product of the individual PDFs, we obtain the combined intrinsic magnification-corrected line fluxes or line flux limits. 
In Table~\ref{tab:obj_comb} we quote the median value and 68\% ($\sim$1$\sigma$) confidence intervals for the final distribution.
Comparing these values with the measurements performed directly on the stack also given in Table~\ref{tab:obj_comb} scaled by the average HFF lens magnification of images A-E ($\langle\mu_\textrm{HFF}\rangle=3.0$), the de-magnified \lya\ and CIV fluxes agree within the errors (especially if the large uncertainties on the magnification are folded into the error budget).
The intrinsic flux limits appear to be roughly a factor of three tighter from the stack than from the Gaussian PDF if the uncertainties on the lens magnification are ignored.
The tentative detection of the OIII]-doublet in the stacks agree with the magnification corrected flux limit set by the Gaussian PDF combination.

\subsection{Interpreting UV Line Fluxes and EWs}
\label{sec:fEWinterp}

As mentioned in Section~\ref{sec:intro}, EWs and flux ratios of rest-frame UV lines are strongly correlated with the ionization strength and the gas-phase metallicity of the stellar populations dominating the SEDs of the emitters \citep[e.g.][]{2016arXiv161003778J,2016MNRAS.462.1757G,2016MNRAS.456.3354F,2015MNRAS.454.1393S,2017MNRAS.464..469S,2017ApJ...836L..14M,2017MNRAS.tmp..190P}.
Hence, the obtained flux ratios (limits) from the GLASS spectra can be used as an indicator of the physical properties of the multiply imaged source behind \RXJ\ as described in the following.

\begin{figure*}
\begin{center}
\includegraphics[width=0.45\textwidth]{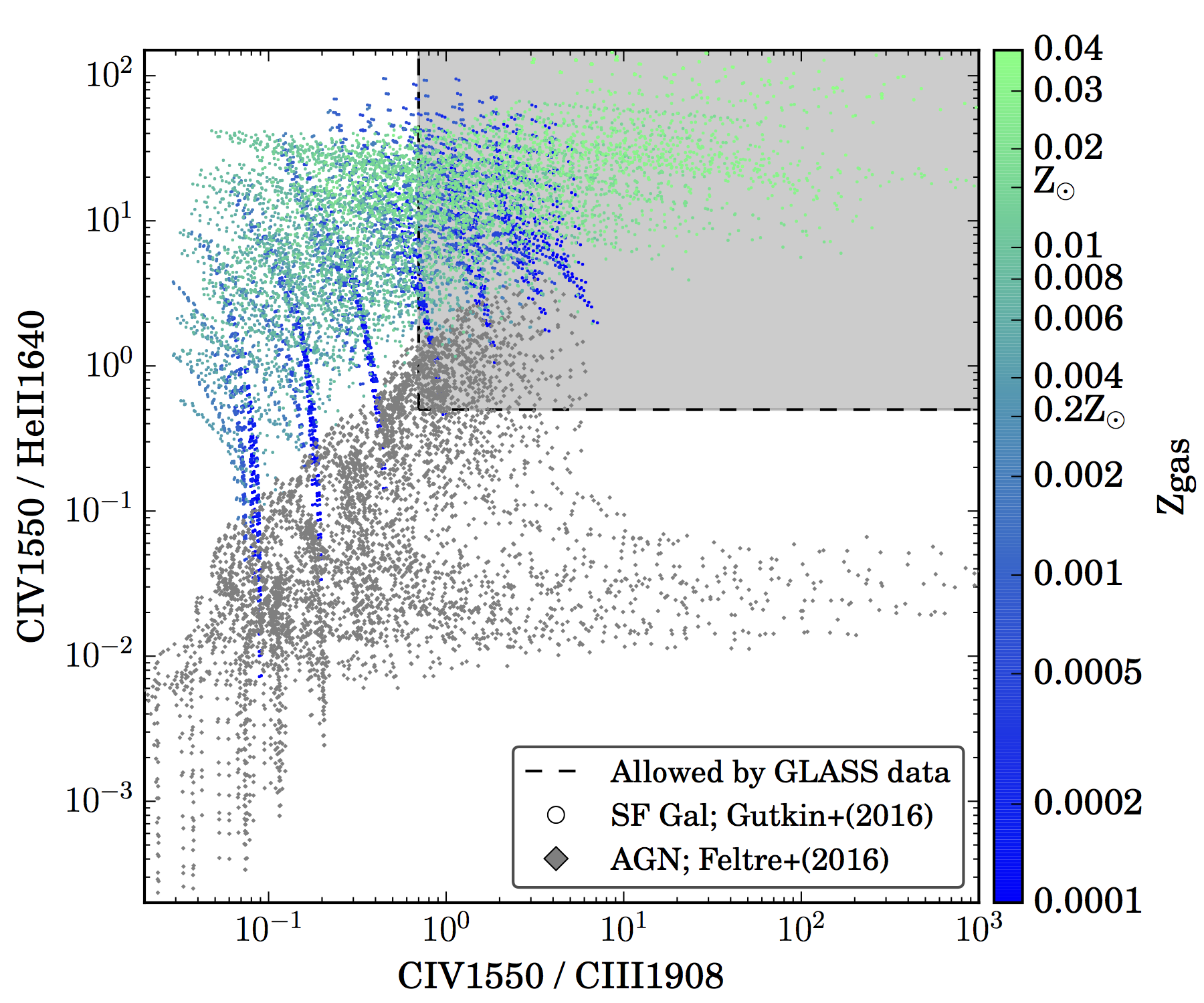}
\includegraphics[width=0.45\textwidth]{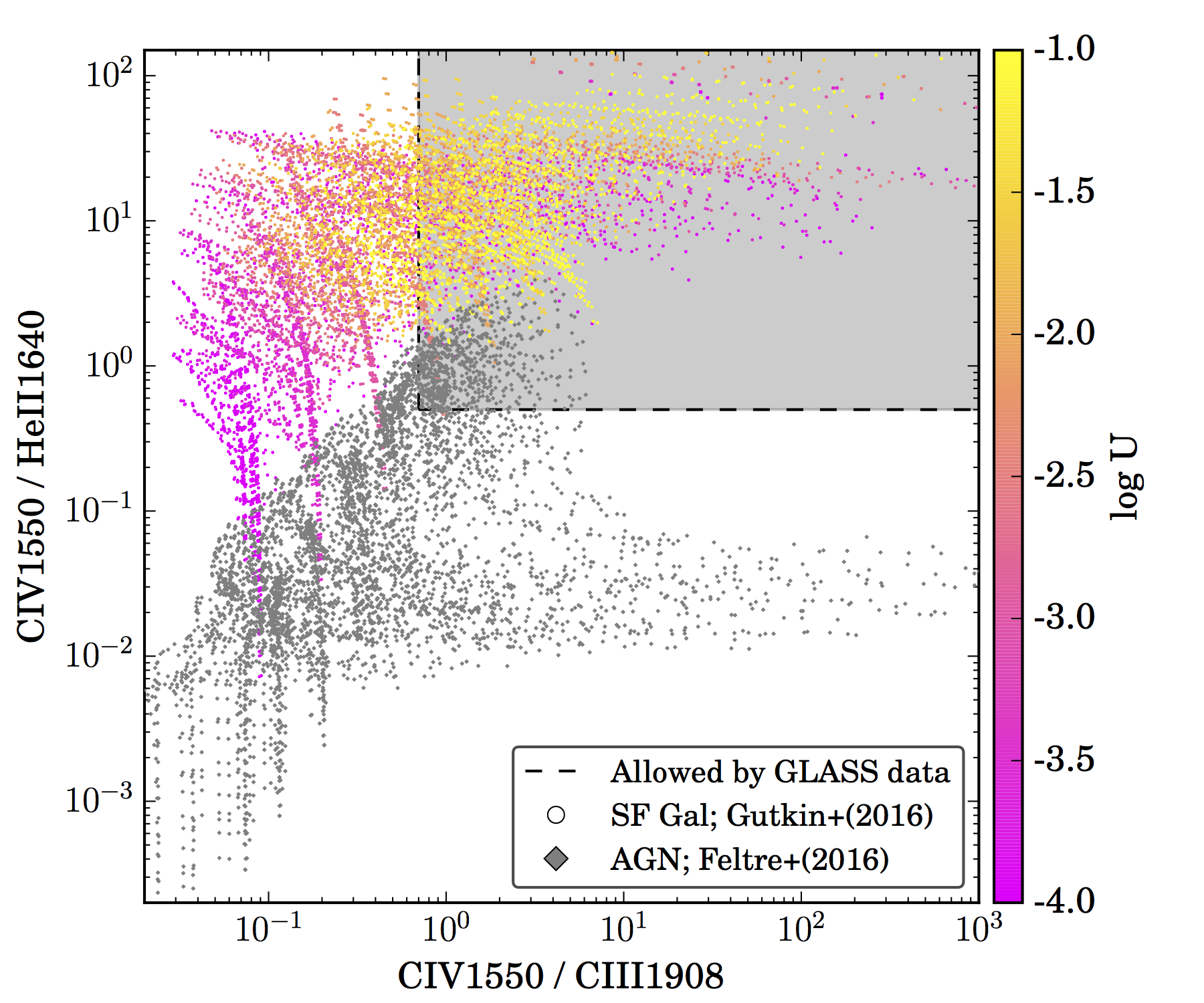}\\
\includegraphics[width=0.45\textwidth]{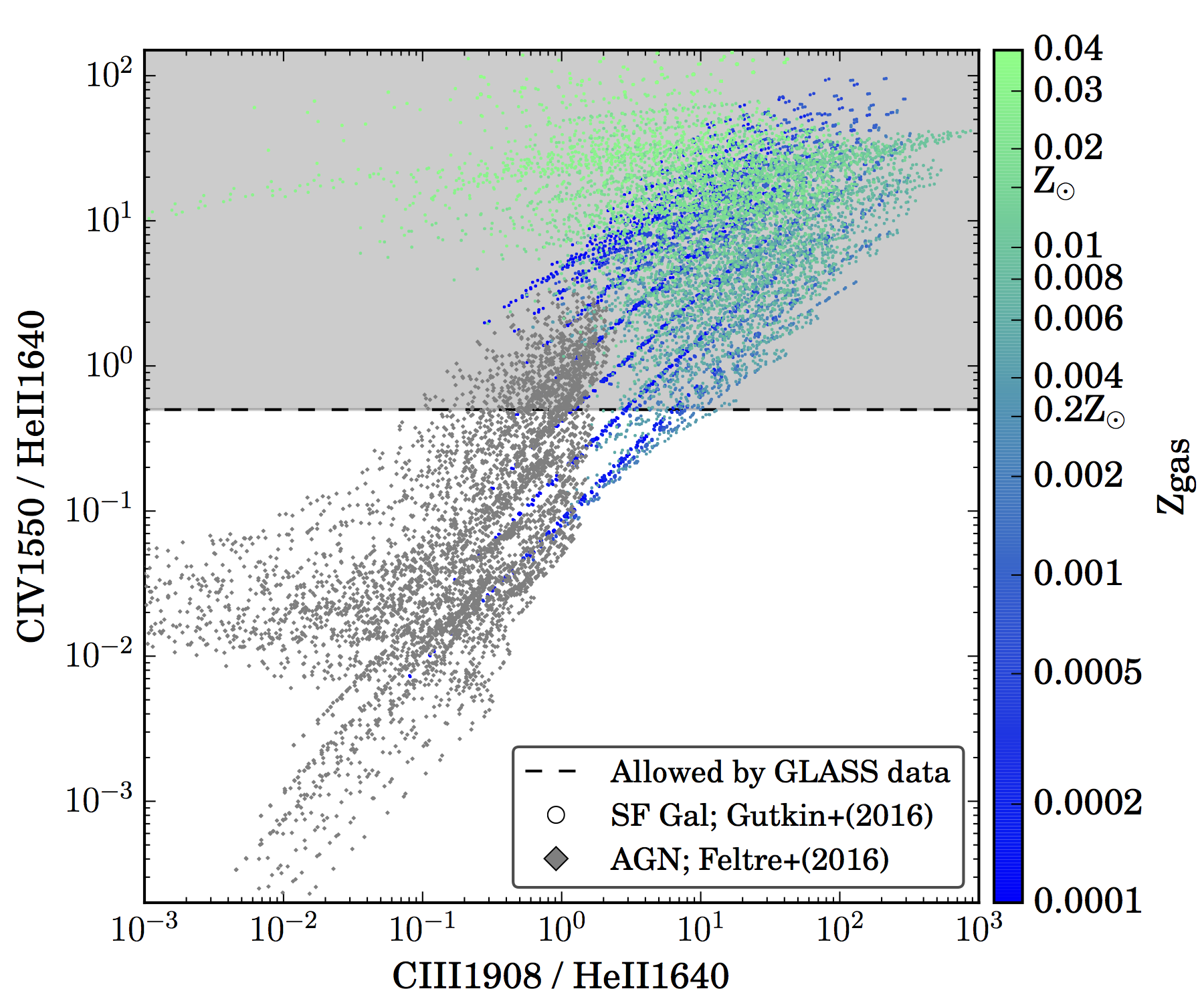}
\includegraphics[width=0.45\textwidth]{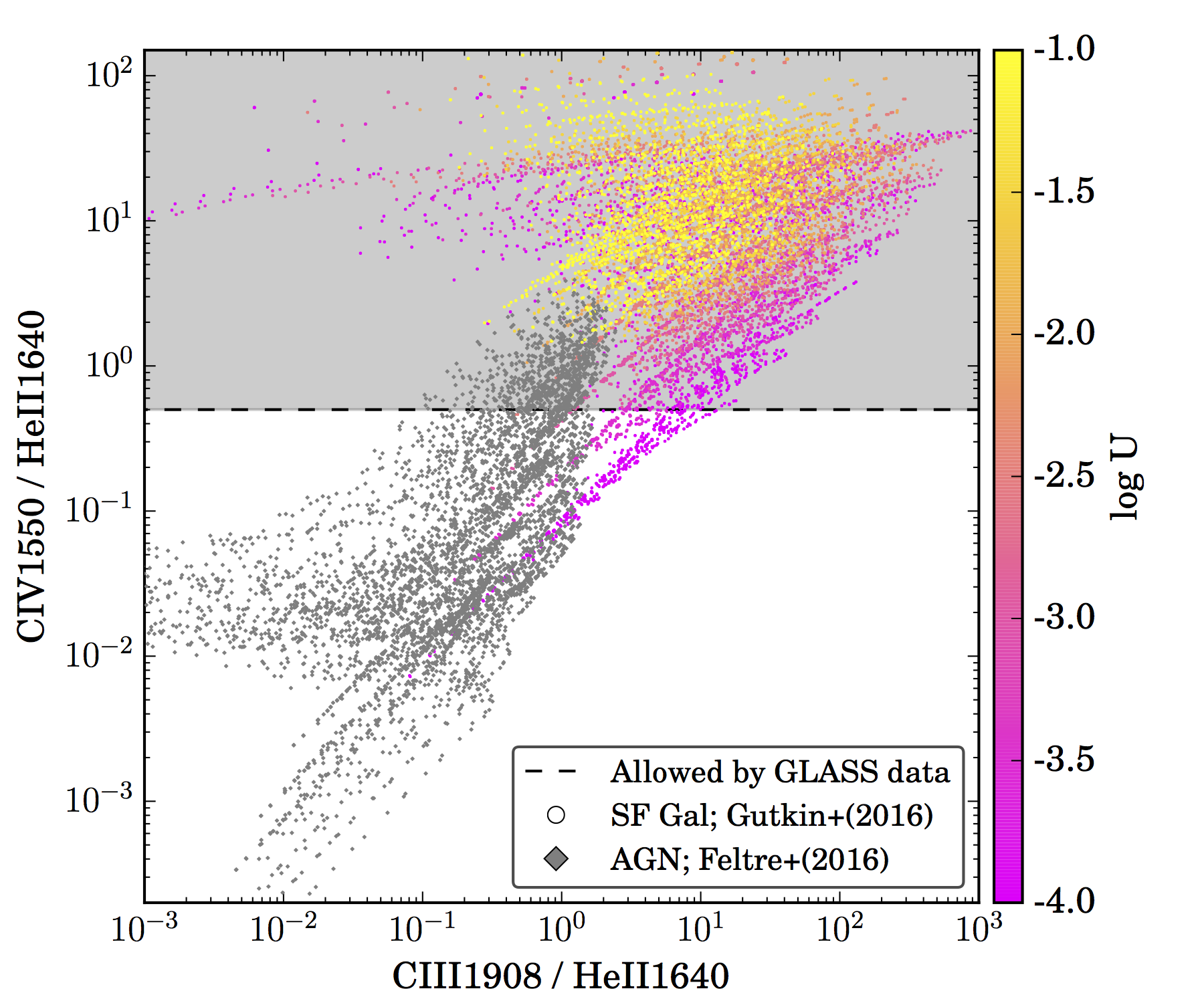}\\
\includegraphics[width=0.45\textwidth]{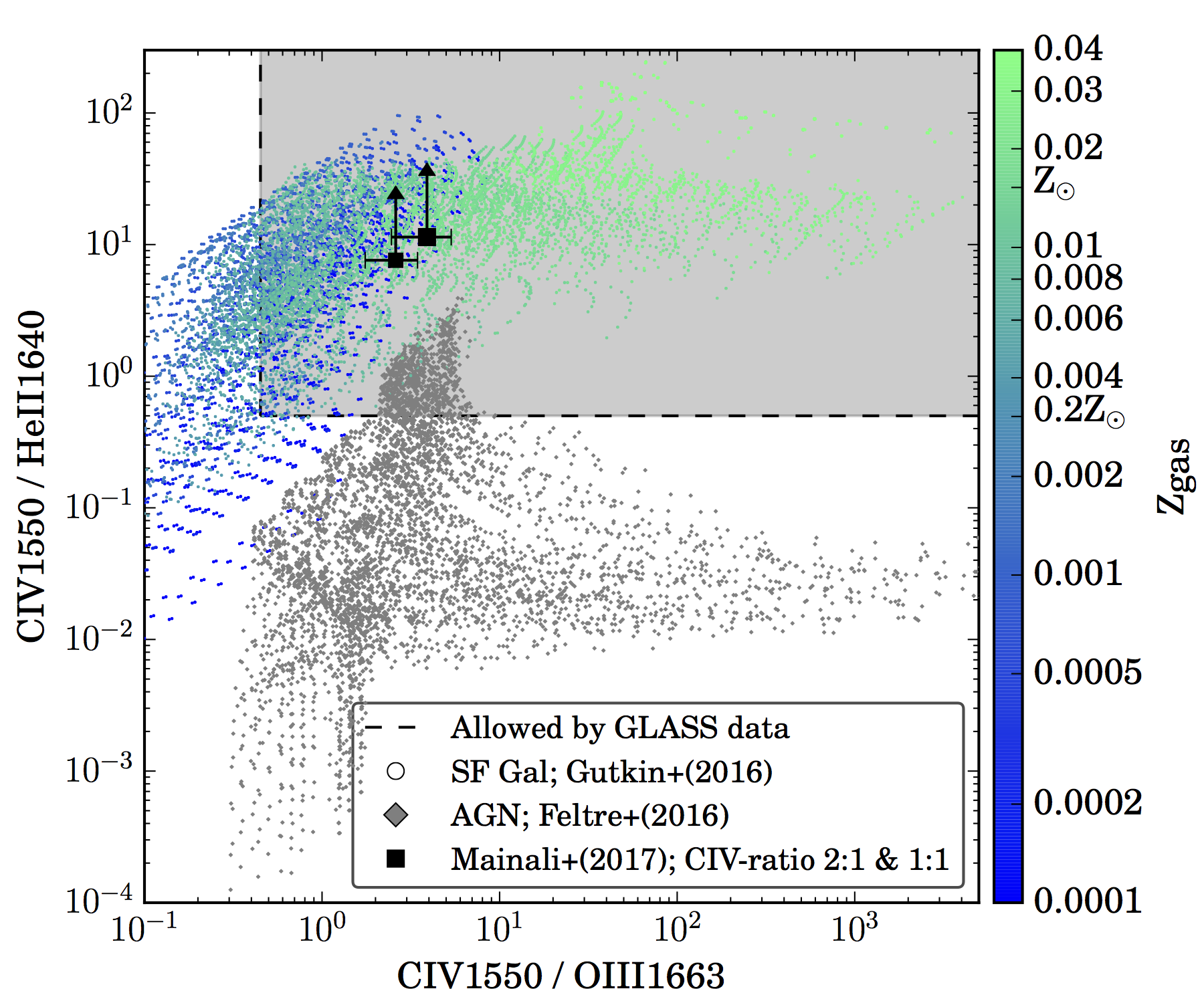}
\includegraphics[width=0.45\textwidth]{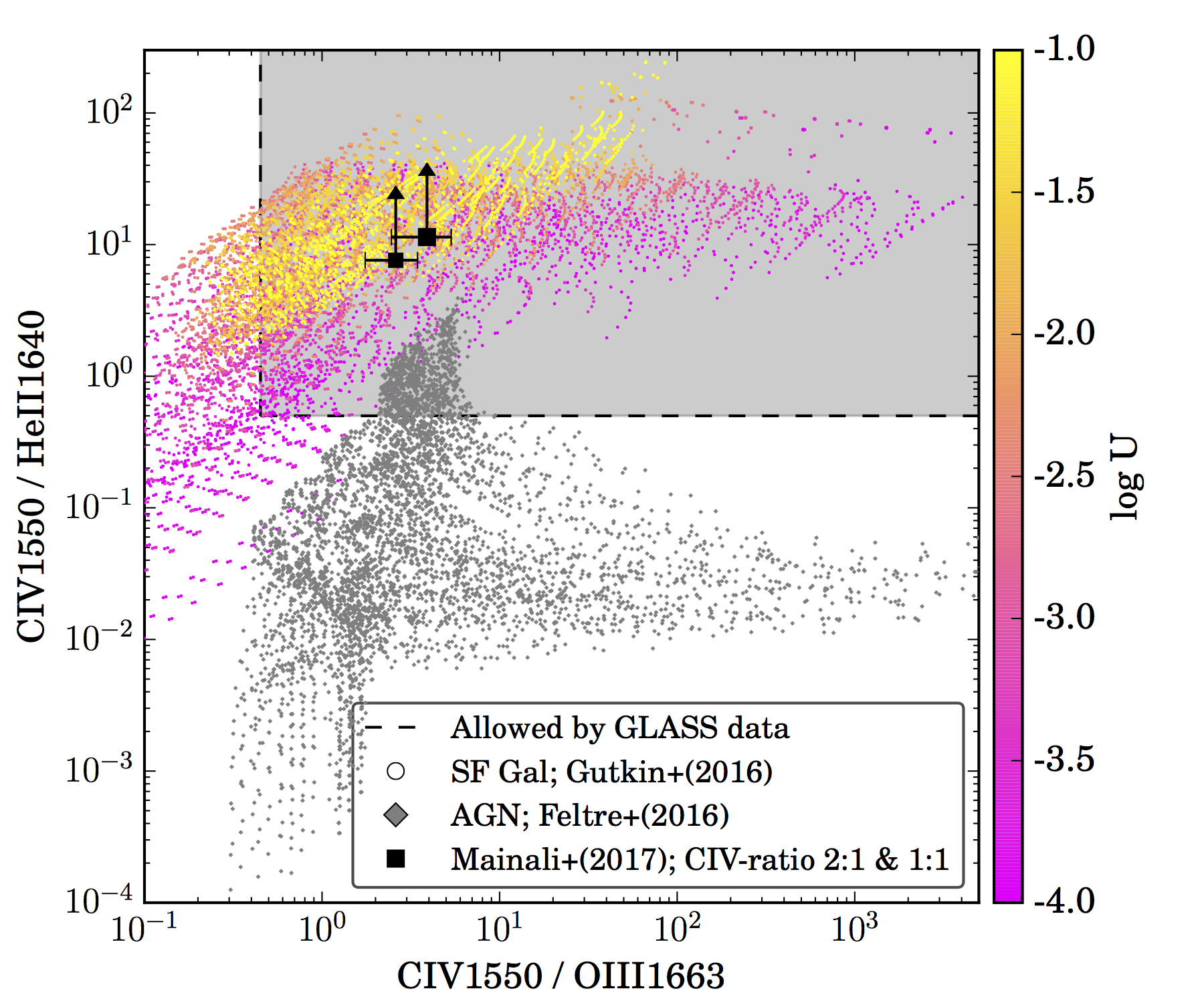}
\caption{
Comparison between the constraints on emission line flux ratios from the GLASS spectroscopy (2$\sigma$ limits; gray shaded regions) of the $z=6.11$ source behind \RXJ, and predictions from recent photoionisation models of the nebular emission from star forming galaxies \citep[colored points;][]{2016MNRAS.462.1757G} and active galactic nuclei \citep[gray diamonds;][]{2016MNRAS.456.3354F}.
`CIV1550', `CIII1908' and `OIII1663' refers to the combined fluxes from the doublets CIV$\lambda\lambda$1548,1551\AA, CIII]$\lambda\lambda$1907,1909\AA, and OIII]$\lambda\lambda$1661,1666\AA, respectively.
All available models from sampling the model input parameters including ionization parameter, gas-phase metallicity, dust-to-metal mass ratios, carbon-to-oxygen ratio, etc. (see \cite{2016MNRAS.462.1757G} and \cite{2016MNRAS.456.3354F} for details) are shown.
The star forming galaxies are color coded according to their gas-phase metallicity ($\textrm{Z}_\textrm{gas}$, left panels) and the ionization parameter ($U$, right panels).
The spectroscopic constraints prefer photoionization models where the CIV emission arise from star formation. 
The bottom panels include the emission line ratios determined for Image E from FIRE spectroscopy (black squares) presented by \citep{2017ApJ...836L..14M}. These estimates agree well with the GLASS limits.
}
\label{fig:ratiomaps}
\end{center}
\end{figure*}

The CIII] EW and the flux ratio between CIV and CIII] are sensitive to the gas-phase metallicity and the ionization parameter of the stellar populations \citep{2016arXiv161003778J,2016MNRAS.462.1757G,2016MNRAS.456.3354F,2017MNRAS.tmp..190P}.
From the GLASS spectra we get rest-frame EW$_\textrm{CIII} \lesssim 20$\AA (1$\sigma$).
As shown by \cite{2015ApJ...814L...6R} EW$_\textrm{CIII}>5$\AA\ is mainly seen in low-metallicity ($<0.5\textrm{Z}_\odot$) systems.
The rest-frame EW$_\textrm{CIII}$ limit from GLASS is in agreement with the EWs of CIII] seen in EW$_\textrm{\lya}\sim70$\AA\ objects in the literature \citep{2015ApJ...814L...6R} 
and the limits presented in recent studies of other sources at the epoch of reionization \citep[e.g.,][]{2015ApJ...805L...7Z,2015Natur.519..327W,2017MNRAS.464..469S}, intermediate redshifts \citep{2017NatAs...1E..52A}, as well as low-$z$ systems \citep{2016arXiv161206866D}. 
The exact levels of CIII] and Z are also dependent on galaxy mass and age of the stellar population.
From the GLASS spectra $\textrm{CIV/CIII}>0.7$ (2$\sigma$).  
\cite{2016arXiv161003778J} stress the rarity of $\textrm{CIV/CIII}>1.0$ (the GLASS data shows a 1$\sigma$ flux ratio of 1.4) illustrating the uncommon properties of this source when compared with the average expected population of UV line emitters in photoionization models.
As we only have upper limits on CIII] and HeII we are currently not in a position to put strong constraints on the UV shock diagnostics of the object using the CIII/HeII vs. CIV/HeII diagram proposed by \cite{1998ApJ...493..571A,1997A&A...323...21V} and explored by \cite{2016arXiv161003778J} and \cite{2016MNRAS.456.3354F}.\footnote{It is worth mentioning that the stellar population synthesis models used to generate these diagnostic diagrams, usually do not incorporate binary massive star systems, which have been found to be important for reproducing observed broad HeII emission at $z\sim2$ \citep{2016ApJ...826..159S}. An example of incorporating the effect of binary stars has been developed as part of the Binary Population and Spectral Synthesis code, \citep[BPASS;][]{2008MNRAS.384.1109E,2009MNRAS.400.1019E,2016MNRAS.462.3302E,2016MNRAS.456..485S} which was used by \cite{2016arXiv161003778J} but not by \cite{2016MNRAS.462.1757G} and \cite{2016MNRAS.456.3354F}.}
However, the GLASS flux limits on CIII] and HeII combined with the CIV detection allow us to assess the likelihood of the ionizing radiation to come from an AGN and to explore the physical properties of the object using the recent photoionization models by \cite{2016MNRAS.456.3354F} and \cite{2016MNRAS.462.1757G}.
In Figure~\ref{fig:ratiomaps} we show the AGN models \citep[gray diamonds,][]{2016MNRAS.456.3354F} and star forming galaxies \citep[colored circles,][]{2016MNRAS.462.1757G} in the parameter spaces spanned by the CIV, CIII], HeII and OIII] emission line ratios.
We have marked the regions allowed by the GLASS flux measurements by the gray shaded regions (using 2$\sigma$ limits). 
From this it is clear that the ionizing emission is most likely from star formation and not generated by AGN as very few of the AGN models fall in the areas allowed by the GLASS data.

The first LAE found to have such extremely high-ionization, but narrow, UV line emission was
MTM095355+545428 \citep[a $z=2.5$ galaxy;][]{1996ApJ...468L...9M}.
Although MTM095355+545428 also has CIV emission substantially
stronger than its CIII]1909, HeII1640, and OIII]1663 lines,
\cite{1996ApJ...468L...9M} concluded that they are primarily photoionized by
a starburst, rather than a Seyfert nucleus.  Their arguments
are similar to the ones made here for the LAE behind \RXJ.

\cite{2015MNRAS.454.1393S} presented an example of a star forming galaxy at $z\sim7$ with $f_\textrm{CIV}/f_\textrm{\lya}>0.2$ similar to CIV-\lya\ flux ratios from hard ionizing narrow-line type II quasar candidates from \cite{2013MNRAS.435.3306A}.
The magnification-corrected combined EW$_\textrm{CIV}=24\pm4$\AA\ and $f_\textrm{CIV}/f_\textrm{\lya}=0.24\pm0.13$ of the $z=6.11$ LAE behind \RXJ\ from the GLASS spectroscopy are very similar to the EW and flux ratio of the object at $z\sim7$ from \cite{2015MNRAS.454.1393S}.
Given the larger rest-frame EW$_\textrm{\lya}=68\pm6$\AA, as for the \cite{2014MNRAS.445.3200S} object, it is unlikely that the unusually high CIV-\lya\ flux ratio is driven mainly by IGM attenuation of the \lya\ line.
Hence, this indicates that the $z=6.11$ source studied here likely has a more extreme ionizing spectrum, similar to those of AGN, than the more modest CIV-\lya\ flux ratios seen in redshift 2 galaxies \citep{2014MNRAS.445.3200S}.

In Figure~\ref{fig:ratiomaps} the star forming models from \cite{2016MNRAS.462.1757G} are color coded according to the gas-phase metallicity (Z, top panels) and the volume averaged ionization parameter ($U$, bottom panels).
Here Z is defined as the mass fraction of all elements heavier than Helium \citep[cf. equation (10) by][]{2016MNRAS.462.1757G}.
In Figure~\ref{fig:ratiomaps} we have marked the solar value Z$_\odot=0.01524$ and 0.2Z$_\odot$ used in the SED modeling described in Section~\ref{sec:SEDs} and quoted as the upper limit for the system by \cite{2014MNRAS.438.1417M}.
Models with a large range of Z and logU are capable of reproducing the GLASS measurements.
However, we note that the relative fraction of Z$\gtrsim0.2$Z$_\odot$ models with logU$\gtrsim-3$ is higher in the regions allowed by the GLASS data.

Flux measurements and flux ratios containing NV measurements are also powerful discriminators when comparing ionizing radiation from star formation with that from AGN.
In Table~\ref{tab:obj} we have listed several potential detections of NV. However, given the S/N of some of these (e.g., Image E PA=053) and the non-detection in other spectra, we only consider these detections as tentative, and therefore avoid drawing conclusions on the NV measurements here.
Also, as the ionization potential of NV is 97.9~eV \citep{NISTAtomicSpectra:uk}\footnote{\url{http://physics.nist.gov/PhysRefData/ASD/ionEnergy.html}} 
and we do not detect any HeII (ionization energy = 54.4~eV), we consider it unlikely for this object to have NV-HeII flux ratios $>1$ given the data. 
However, should this object indeed have a large NV-HeII ratio, this would point towards radiation powered by star formation  from a $\textrm{Z}>0.2\textrm{Z}_\odot$ stellar population, as opposed to an AGN, according to the \cite{2016MNRAS.462.1757G} and \citep{2016MNRAS.456.3354F} models in line with what is seen in Figure~\ref{fig:ratiomaps}.
The spurious nature of the flux excess at the NV location is supported by a non-detection in the MUSE spectra described in Section~\ref{sec:imgF} (even though significant sky-line contamination is present at the NV location in these).

In summary, combining the flux and EW measurements and limits on the rest-frame UV emission lines from the GLASS spectra of the $z=6.11$ LAE behind \RXJ\ draws a picture of non-AGN system with star-formation ionizing radiation arising in a young stellar population (a few -- 200 Myr cf. Section~\ref{sec:SEDs} and \cite{2016arXiv161003778J} models).  

\section{Comparison with Mainali et al. (2017)}
\label{sec:mainali}

\cite{2017ApJ...836L..14M} published a CIV detection in Image~E obtained from 9.17 hours of slit-based spectroscopy with FIRE on Magellan. 
In their work, they extracted the GLASS spectrum of this component.
In the following we will compare their results with the results presented in the current work.

\cite{2017ApJ...836L..14M} measure a \lya\ and CIV$\lambda$1551\AA\ line flux of $3.32\pm0.23\times10^{-17}$erg/s/cm$^2$ and $0.57\pm0.09 \times10^{-17}$erg/s/cm$^2$ from the FIRE spectrum and $1.4\pm0.38\times10^{-17}$erg/s/cm$^2$ and $<0.36 \times10^{-17}$erg/s/cm$^2$ from their extraction of the GLASS spectra of Image E.
The grism flux estimates of \cite{2017ApJ...836L..14M} are roughly 4$\sigma$ smaller for \lya\ than the values presented in Table~\ref{tab:obj}.
This can probably be attributed to the different approaches in reduction and potential over/under-subtraction of contamination from the spectra before fluxes are measured.
The estimated EW$_\textrm{\lya}$ from the GLASS spectra also disagree by $\sim$4$\sigma$ indicating that the estimated continuum fluxes are similar. 
Here we used the F105W broadband magnitude from our ICL-subtracted photometry whereas \cite{2017ApJ...836L..14M} estimated the continuum from SED fitting to their own \HST\ and Spitzer/IRAC photometry.
Comparing to the EW estimated from the FIRE spectrum, the combined EW presented in Table~\ref{tab:obj} differs by 3.6$\sigma$. 
Furthermore, comparing to the EW$_\textrm{\lya}$ estimated by \cite{2013A&A...559L...9B} of $79\pm10$\AA\ our estimate agrees with theirs at the 3.0$\sigma$ level.
Hence, our space-based EW$_\textrm{\lya}$ estimate is bracketed by the FIRE and \cite{2013A&A...559L...9B} ground-based slit spectroscopic estimates, where correctly accounting for slit-losses is always challenging.
 
The independent CIV detection and CIII] flux limit from the FIRE and grism spectra, respectively, presented by \cite{2017ApJ...836L..14M} are in good agreement with the measurements presented in Table~\ref{tab:obj}.
Following the example of \cite{2017ApJ...836L..14M} and assuming that the flux line ratio $f_\textrm{CIV$\lambda$1548}/f_\textrm{CIV$\lambda$1551}=1$ for  metal poor CIV emitters, the unresolved CIV doublet should have flux of $\sim1.14\times10^{-17}$erg/s/cm$^2$ in the GLASS spectra.
This is within 3$\sigma$ of the values we measure in the spectra of image E.
For a 1:1 flux ratio the EW$_\textrm{CIV}$ from \cite{2017ApJ...836L..14M} is in good agreement ($\sim$1$\sigma$) with the combined EW$_\textrm{CIV}=24\pm4$\AA\ from the GLASS spectra presented in Table~\ref{tab:obj_comb}.
The theoretically expected ratio of the two components in the CIV doublet is 2:1 \citep[e.g.,][]{1983A&A...122..335F} which has been observed in young low-metallicity objects at $z=3.1$ presented by \cite{2016ApJ...821L..27V}.
With a CIV doublet flux ratio of two the \cite{2017ApJ...836L..14M} detection would result in a combined CIV flux of $\sim1.71\times10^{-17}$erg/s/cm$^2$, which is within 2.4$\sigma$ of the GLASS measurements.

At the location of NV we see a flux excess in the GLASS spectra analyzed in this study.
However, as mentioned previously the brightness in individual spectra (including the extracted spectra of Image E) and lack of detection in other spectra, leads us to believe that this feature is spurious.
Hence, on the NV flux measurements, we can only conclude that our combined limit is likely biased by spurious detections, but that the individual flux limits do not dis-agree with the limit obtained from the FIRE spectrum by \cite{2017ApJ...836L..14M}.

The GLASS limits on HeII presented here agrees with the limit obtained from the FIRE spectrum ($<0.15\times10^{-17}$erg/s/cm$^2$) presented by \cite{2017ApJ...836L..14M}.

Finally, we put observed upper limits on the flux (EW) of OIII] of 
$<0.85\times10^{-17}$erg/s/cm$^2$ ($<23$\AA) and $<0.70\times10^{-17}$erg/s/cm$^2$ ($<19$\AA)
from the GLASS spectra from the two PAs of image E, respectively (cf. Table~\ref{tab:obj}).
These limits are in agreement with the estimates from the FIRE spectrum. 
However, given that the OIII] doublet is unresolved in the GLASS spectra, the combined flux of the two components of the OIII] doublet from the FIRE spectrum ($0.44\times10^{-17}$erg/s/cm$^2$) should in principle be detectable in the stack of the GLASS spectra from all multiply imaged components.
The flux excess at the OIII] wavelength seen in the GLASS stack in Figure~\ref{fig:stack} is as mentioned formally a 3$\sigma$ detection even though we consider it only tentative (see Section~\ref{sec:fcorrection} and Table~\ref{tab:obj_comb}).
Nevertheless, we do note that the measured observed flux in the GLASS stack is in agreement with the FIRE detection reported by \cite{2017ApJ...836L..14M}. 

In the bottom panels of Figure~\ref{fig:ratiomaps} we have over-plotted the FIRE emission line flux ratios of CIV, HeII and OIII from \cite{2017ApJ...836L..14M} (assuming a CIV doublet line ratio of both 2:1 and 1:1).
These measurements are in good agreement with the GLASS limits presented in the current study. 

Based on the detections and upper limits presented by \cite{2017ApJ...836L..14M} they arrive at conclusions similar to the ones presented in Section~\ref{sec:fEWinterp}, i.e. that the object has a hard ionizing spectrum arising from star formation rather than AGN activity.
They also stress that the system agrees with low-metallicity stellar populations in good agreement with the metallicity indicated by the GLASS data presented here.
However, as described in Section~\ref{sec:fEWinterp} and shown in Figure~\ref{fig:ratiomaps}, also higher metallicities sampled in current photoionization modeling \citep{2016MNRAS.462.1757G,2016arXiv161003778J,2016MNRAS.456.3354F} can reproduce the current measurements.

\section{Conclusions}
\label{sec:conc}

We have presented the detection of CIV$\lambda\lambda$1548,1551\AA\ (and \lya) in a known quintuply imaged object \citep{2014MNRAS.438.1417M} at $z=6.11$ behind the Hubble Frontier Fields cluster \RXJ\ (Abell S1063).
The emission was detected in \HST\ slitless grism spectroscopy obtained as part of the GLASS observations.
We put informative upper limits on other rest-frame UV emission lines including the CIII]$\lambda\lambda$1907,1909\AA\ and OIII]$\lambda\lambda$1661,1666\AA\ doublets and HeII$\lambda$1640\AA.
By combining these measurements with updated ICL-subtracted photometry on all components we have drawn the following main conclusions from the data:

\begin{enumerate}

\item We detect CIV$\lambda\lambda$1548,1551\AA\ at more than 3$\sigma$ in two of the five components of the multiply imaged $z=6.11$ LAE behind \RXJ. In two further components we see marginal CIV emission at the 2$\sigma$ level.
A marginal flux excess of the previously published OIII]$\lambda\lambda$1661,1666\AA\ emission \citep{2017ApJ...836L..14M} is seen in the stack of all components of the system. 

\item The equivalent widths combined from all components of the multiply imaged system of CIII] (EW$_\textrm{CIII]} < 20$\AA), CIV (EW$_\textrm{CIV}=24^{+4}_{-4}$\AA) and \lya\ (EW$_\textrm{\lya} = 68^{+6}_{-6}$) are in good agreement with values published in the literature for this system, and agree well with expectations from previous measurements of similar objects.

\item The flux and EW measurements and limits of the rest-frame UV lines support a picture of a high-redshift galaxy with ionization produced by rapid star formation from a young stellar population ($<50$Myr old). 
The presented data do not favor an AGN-supported radiation field.

\item Our results and conclusions are in good agreement with the independent detection of CIV (and OIII]$\lambda\lambda$1661,1666\AA) in image E of the system by \cite{2017ApJ...836L..14M}.

\item The spatial extent of the \lya\ and CIV emission is un-resolved in the GLASS spectra, i.e. similar to the spatial extent of the PSF estimated from spectra of stars. Hence, we are unable to analyze the potential difference in the spatial extent of the regions emitting in \lya\ and CIV with the current data.

\item Based on GLASS and MUSE spectroscopy, the HFF cluster lens models, as well as the updated photometric measurements presented in the current study, we confirm that the potential sixth image proposed by \cite{2015A&A...574A..11K} (image F), is more likely a low-$z$ [OII]$\lambda\lambda$3726,3729\AA\ emitter or another LAE at $z\sim6.1$.
\end{enumerate}

Hence, the quintuply imaged LAE at $z=6.11$ studied here is another example of a highly ionizing source powered by star formation in the first 900Myr after Big Bang.
Similar sources have likely had an important effect on the ionization history of the Universe.
To quantify this effect, larger samples of similar sources need to be assembled to assess whether such galaxies are frequently occurring at high redshift, or have only been discovered in the recent literature due to their extreme properties.
JWST will help explore this domain by enabling detection of larger, potentially more typical, populations of galaxies at the EoR  ($z>6$) through imaging and spectroscopy in the rest-frame optical.

\acknowledgments

\footnotesize
We would like to thank A. Fontana, M. Castellano. E. Merlin and the ASTRODEEP collaboration for sharing their photometric pipeline and assisting us in developing the photometric catalogs on \RXJ.
We would like to thank E. Vanzella for valuable discussions and input.
T.M. acknowledges support from the Japan Society for the Promotion of Science (JSPS) through JSPS research fellowships for Young Scientists.
T.J. acknowledges support provided by NASA through Program \# HST-HF2-51359 through a grant from the Space Telescope Science Institute, which is operated by the Association of Universities for Research in Astronomy, Inc., under NASA contract NAS 5-26555.
%
This paper is based on observations made with the NASA/ESA Hubble
Space Telescope, obtained at STScI,
on observations made with the Spitzer Space Telescope, which is operated by the Jet Propulsion Laboratory, California Institute of Technology under a contract with NASA,
and on data products from observations made with ESO Telescopes at the La Silla Paranal Observatory under programme ID 60.A-9345(A).
%
We acknowledge support through
grants HST-13459, HST-GO13177 and HST-AR13235.
STScI is operated by AURA, Inc. under NASA contract NAS 5-26555. 
B.V. acknowledges the support from an Australian Research Council Discovery Early Career Researcher Award (PD0028506)
This research made use of the following open-source packages for Python and we are thankful to the  developers of these:
Astropy \citep{2013A&A...558A..33A},
APLpy \citep{2012ascl.soft08017R},
iPython \citep{Perez:2007hy},
numpy \citep{vanderWalt:2011dp},
matplotlib \citep{Hunter:2007ux}, 
and
PyFITS which is a product of the Space Telescope Science Institute, which is operated by AURA for NASA.
%
The lens models were obtained via the HFF page on the Mikulski Archive for Space Telescopes (MAST).

\normalsize
\begin{appendix}
\section{The Individual GLASS Spectra of the Multiple Images}
\label{app:individualspec}

\begin{figure*}
\begin{center}
\includegraphics[width=0.95\textwidth]{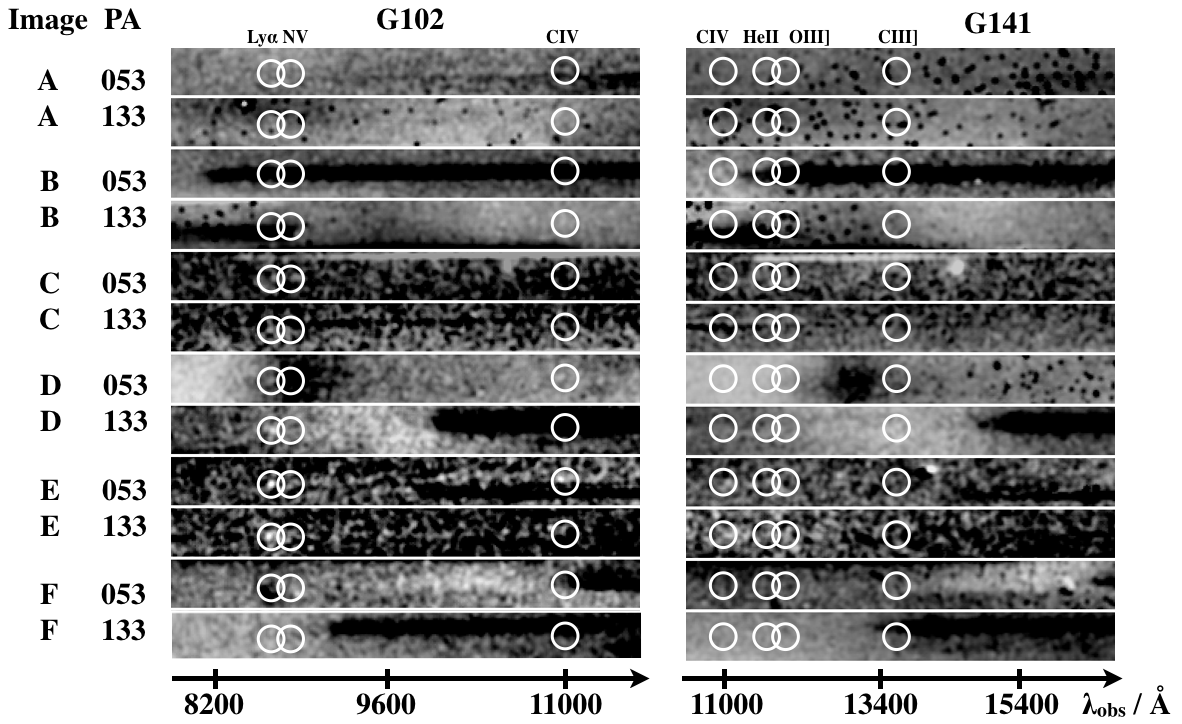}
\caption{Overview of the individual GLASS spectra of all six proposed components of the system (Image A--F as listed on the left). 
Each source image has a spectrum taken in G102  (left) and G141 (right) at both PAs as indicated.
The white circles mark the expected location of the rest-frame UV emission lines \lya, NV, CIV, HeII, OIII] and CIII].
The circles are 200\AA$\times$1$\farcs$1 and 400\AA$\times$1$\farcs$1 in the observed frame for G102 and G141, respectively.
}
\label{fig:individualspec}
\end{center}
\end{figure*}

In Figure~\ref{fig:individualspec} we show the 24 individual GLASS spectra of the six proposed components (Images A--F) of the LAE emitter at $z=6.11$ behind \RXJ. Each spectrum has been subtracted the contamination model and has been scaled to highlight the emission line flux excesses marked by the white circles.
The stack cutouts shown in Figure~\ref{fig:stack} have been generated by combining these spectra excluding the spectra for image F (which is likely not part of the system cf. Section~\ref{sec:imgF}) and Image B at PA=053 (due to the large contamination at this PA).

\end{appendix}


\end{document}